\documentclass[twocolumn,notitlepage]{aastex61}

\usepackage{hyperref}
\usepackage{latexsym}
\usepackage{graphicx,epsf,epsfig}
\usepackage{amssymb,amsmath}
\usepackage{cancel}
\usepackage{soul}

\defcitealias{2017ApJ...846..125B}{Paper~I}

\graphicspath{{./figures/}}

\submitjournal{ApJ}

\begin{document}
\title{The discreteness-driven relaxation of collisionless gravitating
  systems: entropy evolution in external potentials, N-dependence and
  the role of chaos}

\author{Leandro {Beraldo e Silva}} \affiliation{Universidade de S\~ao
  Paulo, Instituto de Astronomia, Geof\'isica e Ci\^encias
  Atmosf\'ericas, Departamento de Astronomia, CEP 05508-090, S\~ao
  Paulo, SP, Brasil} \affiliation{Department of Astronomy, University
  of Michigan, Ann Arbor, MI 48109, USA} \email{lbs@usp.br}
\author{Walter {de Siqueira Pedra}} \affiliation{Universidade de S\~ao
  Paulo, Instituto de F\'isica, Departamento de F\'{\i}sica
  Matem\'atica, CP 66318, CEP 05314-970, S\~ao Paulo, SP, Brasil}
\author{Monica Valluri} \affiliation{Department of Astronomy,
  University of Michigan, Ann Arbor, MI 48109, USA} \author{Laerte
  Sodr\'e}\affiliation{Universidade de S\~ao Paulo, Instituto de
  Astronomia, Geof\'isica e Ci\^encias Atmosf\'ericas, Departamento de
  Astronomia, CEP 05508-090, S\~ao Paulo, SP, Brasil}
\author{Jean-Bernard Bru}\affiliation{Departamento de Matem\'aticas,
  Facultad de Ciencia y Tecnolog\'ia, Universidad del Pa\'is Vasco,
  Apartado 644, 48080 Bilbao, Spain} \affiliation{BCAM - Basque Center
  for Applied Mathematics, Mazarredo, 14, 48009 Bilbao, Spain}
\affiliation{IKERBASQUE, Basque Foundation for Science, 48011 Bilbao,
  Spain}

\begin{abstract}
  We investigate the old problem of the fast relaxation of
  collisionless $N$-body systems which are collapsing or perturbed,
  emphasizing the importance of (non-collisional) discreteness
  effects. We integrate orbit ensembles in fixed potentials,
  estimating the entropy to analyze the time evolution of the
  distribution function. These estimates capture the correct physical
  behavior expected from the 2nd Law of Thermodynamics, without any
  spurious entropy production. For self-consistent (i.e. stationary)
  samples, the entropy is conserved, while for non-self-consistent
  samples, it increases within a few dynamical times, stabilizing at a
  maximum (even in integrable potentials). Our results make
  transparent that the main ingredient for this fast collisionless
  relaxation is the discreteness (finite $N$) of gravitational systems
  in any potential. Additionally, in non-integrable potentials, the
  presence of chaotic orbits accelerates the entropy
  production. Contrary to the traditional violent relaxation scenario,
  our results indicate that a time-dependent potential is not
  necessary for this relaxation. For the first time, in connection
  with the Nyquist-Shannon theorem we derive the typical timescale
  $T/\tau_{cr}\approx 0.1 N^{1/6}$ for this discreteness-driven
  relaxation, with slightly weaker $N$-dependencies for non-integrable
  potentials with substantial fractions of chaotic orbits. This
  timescale is much smaller than the collisional relaxation time even
  for small-$N$ systems such as open clusters and represents an upper
  limit for the relaxation time of real $N$-body collisionless
  systems. Additionally, our results reinforce the conclusion of
  Beraldo e Silva et al. (2017) that the Vlasov equation does not
  provide an adequate kinetic description of the fast relaxation of
  collapsing collisionless $N$-body systems.
\end{abstract}

\section{Introduction}
\label{sec:introduction}
The relaxation of self-gravitating systems such as globular clusters,
galaxies and dark matter halos is characterized by at least two time
scales: the crossing (dynamical) timescale
$\tau_{cr}\approx R/\langle v\rangle$ and the typically much longer
collisional relaxation timescale
${\tau_{col}\approx (N/\ln N) \tau_{cr}}$, where $R$ is the system's
size, $\langle v \rangle$ is a characteristic velocity and $N$ is the
number of bodies -- see \cite{Binney_2008}.

For small timescales in comparison to $\tau_{col}$, the collisional
relaxation can be neglected, and the system is said to be
collisionless. Despite being collisionless, these systems still
undergo an initial phase of fast relaxation (in a few dynamical time
scales), coined violent relaxation in the scenario proposed by
\cite{LyndenBell_1967}. In this scenario, the key ingredient is a
time-varying collective potential, which changes individual
energies. These indirect energy exchanges among the stars (through the
time-varying potential) are further interpreted as driving the
relaxation.

Most theoreticians discuss this fast collisionless relaxation in terms
of infinitely divisible distribution functions, i.e. in the limit
$N\rightarrow \infty$, despite the fact that real stellar systems are
composed of a finite $N$ and are not infinitely divisible and
smooth. In this context, the kinetic evolution of collisionless
systems is traditionally expected to be described by the
Vlasov-Poisson equation (with discreteness effects only producing
corrections in the long-term evolution through collisional
relaxation):
\begin{equation}
\label{eq:vlasov}
  \frac{df}{dt} \equiv \frac{\partial f}{\partial t} +
  \vec{v}\cdot\frac{\partial f}{\partial \vec{r}} -
  \frac{\partial \phi}{\partial \vec{r}}\cdot\frac{\partial f}{\partial \vec{v}} = 0,
\end{equation}
where $f(\vec{r},\vec{v},t)$ is the distribution function,
representing the probability of finding a test particle in position
$\vec{r}$ and velocity $\vec{v}$. In Eq.~\eqref{eq:vlasov},
$\phi(\vec{r},t)$ is the collective gravitational potential,
considered as an external potential for the test particle and
self-consistently related to the distribution function by means of the
Poisson equation
\begin{equation}
\label{eq:poisson}
  \nabla^2\phi = 4\pi G \int d^3\vec{v}\,f(\vec{r},\vec{v},t).
\end{equation}

Eq.~\eqref{eq:vlasov} yields an evolution for the distribution
function $f$ which is time reversible and conserves the entropy
$S$. In fact, as shown by \cite{Tremaine_Henon_Lynden_Bell_1986},
defining
\begin{equation}
  \label{eq:S_def}
  S(t) \equiv -\int f\ln f \,d^3\vec{r}\,d^3\vec{v},
\end{equation}
then if ${df/dt = 0\Rightarrow dS/dt = 0}$.

\cite{2017ApJ...846..125B} (hereafter {\bf Paper I}) ran N-body
simulations and studied the evolution of the entropy defined by
Eq.~\eqref{eq:S_def}. This quantity was estimated at each snapshot
with well established mathematical prescriptions, and this estimate
was shown to have a fast increase in the early stages and a slow,
$N$-dependent and almost linear increase in the long-term
evolution. This long-term behavior is well described by the
orbit-averaged Fokker-Planck equation modelling the collisional
relaxation. On the other hand, in \citetalias{2017ApJ...846..125B} it
is argued that, since the Vlasov-Poisson equation implies entropy
conservation, the observed entropy increase on a dynamical timescale
is an indication of non-validity of the Vlasov-Poisson equation in the
violent relaxation.

While there is still no mathematically rigorous proof of the
Vlasov-Poisson equation in the full self-gravitating N-body problem,
the Vlasov equation has already been proven for any fixed
\textit{finite} time interval and (initially) independently
distributed particles subject to an external potential with integrable
spatial gradient and bounded interparticle forces, in the limit
$N\rightarrow \infty$ -- see \cite{Dobr}. See also
\citetalias{2017ApJ...846..125B} for a summary of recent mathematical
results. Notice that the term ``Vlasov-Poisson equation'' is used for
the special case of a Coulomb two-body potential in self-gravitating
systems (Eq.~\eqref{eq:poisson} being the corresponding
self-consistency condition), whereas ``Vlasov equation'' refers to
generic potentials.

In the current work, instead of focusing on the evolution of a
self-gravitating $N$-body system, we study the simpler problem of
ensembles of orbits interacting only with fixed external potentials,
chosen on the basis of their phase-space properties. As we show below,
the entropy evolution in these potentials agrees with what is expected
from the 2nd law of Thermodynamics, i.e. it increases whenever the
initial state does not correspond to an invariant state compatible
with the macroscopic constraints. On the other hand, the entropy
evolution of non-self-interacting particles in external potentials
raises questions regarding the non validity of the Vlasov equation in
a situation where it was expected to be valid. In this work we show
that, even though the Vlasov-Poisson equation might be valid in the
limit $N\rightarrow \infty$, its use in the description of the
evolution of gravitational systems can be problematic for three,
correlated, reasons: (1) on physically relevant timescales, the
convergence of the discrete, real problem to the continuous limit
$N\rightarrow \infty$ can be too slow (i.e. with too weak
$N$-dependence), invalidating the use of this continuous limit for
timescales of interest and values of $N$ that are typical of real
systems; (2) previous rigorous results discussing the validity of the
Vlasov equations in the large $N$ limit are not uniform with respect
to time \citep[typically, the timescale for which the Vlasov equation
is proven to be valid grows only as $\ln N$ -- see][]{BoePick,
  LaPick}, suggesting that \textit{in general} the Vlasov equation
fails to describe the macroscopic state of finite $N$ systems
(initially in non-stationary states) for sufficiently large times; (3)
its physical content is equivalent to the Newtonian description
(i.e. at the microscopic level of single trajectories), lacking the
physical content necessary to describe emergent phenomena, i.e. on a
macroscopic level, particularly the time irreversibility, as expected
from a kinetic equation -- see \S~\ref{sec:micro_vs_macro} and
\S~\ref{sec:meaning_vlasov}.

The orbit integration in fixed external potentials carried out in this
work allows us to critically revise some of the conclusions drawn in
\citetalias{2017ApJ...846..125B} and to explore in more detail the
meaning of the entropy estimators and their relation to the Vlasov
equation. This analysis also sheds some light on the fast (violent)
collisionless relaxation of $N$-body systems and its time
irreversibility. Further, by comparing results obtained for ensembles
of orbits in integrable and non-integrable potentials we investigate
the role of chaos on the entropy evolution.

The current results supplement those obtained in
\citetalias{2017ApJ...846..125B}, highlighting the origin of the
observed rapid entropy increase. Additionally, we confirm the absence
of any artificial, non-physical, inputs in
\citetalias{2017ApJ...846..125B}, which might have been introduced by
the numerical methods used. In particular, we answer the following
questions:
\begin{enumerate}
\item{Why is there time irreversibility in the fast relaxation of
    collisionless systems (\citetalias{2017ApJ...846..125B}) although
    the equations of motion for individual trajectories are
    time-reversible?}
\item{Does the entropy increase observed in
    \citetalias{2017ApJ...846..125B} result from truncation errors
    always involved in the integration of orbits on a computer?}
\item{Was the observed entropy increase due to artificial correlations
    introduced by the entropy estimators?}
\item{Was the observed entropy increase a result of information loss
    due to coarse-graining when estimating the entropy?}
\end{enumerate}

In \S~\ref{sec:micro_vs_macro}, we discuss the concept of time
irreversibility, making explicit the differences between the
descriptions on the micro and macroscopic levels and thereby answer
question 1 above. In \S~\ref{sec:violent_relax} we summarize possible
ingredients for the time irreversibility of the early collisionless
relaxation. In \S~\ref{sec:estimators}, we introduce the entropy
estimator and show its quantitative agreement with the theoretical
expression in simple cases. In \S~\ref{sec:harmonic} we start the
study of the dynamical evolution of gravitational systems with the
harmonic potential, showing that the numerical scheme is able to
recover macroscopic time reversibility when it is present. In
\S~\ref{sec:plummer} we integrate ensembles of orbits in the Plummer
potential, showing that the entropy estimates behave in agreement with
the 2nd law of Thermodynamics, i.e. increasing when the ensemble is
far from an invariant distribution and being conserved for an
invariant one. These two sections answer questions 2 and 3 above. The
role of chaos for time irreversibility is discussed in
\S~\ref{sec:ellipsoid}, where we integrate ensembles of orbits in an
ellipsoidal model, studying the $N$-dependence of the entropy
evolution. We also perform frequency analysis for these orbits in
order to investigate the phase-space structure for the underlying
potential models, allowing us to also estimate the fractions of
regular and chaotic orbits in each model. Finally, in
\S~\ref{sec:meaning_estimator} and \S~\ref{sec:meaning_vlasov} we
discuss the meaning of the entropy estimator and its relation to the
Vlasov equation, answering criticism 4 above. We conclude in
\S~\ref{sec:conclusions}.

\section{Micro vs. macroscopic description}
\label{sec:micro_vs_macro}
Given a system composed of $N$ particles, Classical Mechanics presents
a microscopic characterization of its evolution, describing the motion
of every single particle in the system, by means of time-reversible
equations. Since this generally involves a huge number ($3N$) of
degrees of freedom, the standard strategy is to reduce this number, by
studying quantities characterizing the system as a whole, e.g. total
energy, the virial ratio, pressure, temperature, entropy and so on.

The theory \textit{par excellence} to study such global quantities is
Thermodynamics. The 2nd law of Thermodynamics expresses the fact that
natural phenomena are generally time irreversible at a
\emph{macroscopic} level, and are parametrized by a quantity which
increases with time, the entropy. This agrees with observed phenomena,
like diffusion of ink in water or evaporation of a perfume in a room.

Let us emphasize that the distinction between micro and macroscopic
levels does not refer to the system size. Instead it refers to the
kind of description to be made: description of the movement of each
constituent element on the one hand, versus the description of the
system as a whole on the other hand. Thermodynamics refers to the
macroscopic level and is independent of any specific theory used to
model the microscopic behavior. This is illustrated by the fact that
it has remained practically intact in the transition from Classical to
Quantum and Relativistic Mechanics, and also in face of the new
paradigm associated with chaos.

At the same time, when possible, the interpretation of macroscopic
phenomena in terms of a mechanical description for individual
particles can be illuminating. However, when this is difficult or
impossible, it does not mean that the macroscopic effect is not real,
but it makes clear the difficulties and limitations of the
reductionist point of view: phenomena that occur at the macroscopic
level generally need to be seen as new phenomena, and not as simple
collective manifestations of the microscopic phenomena. In
\cite{Anderson393}'s words: ``more is different'' \citep[see
also][]{Uhlenbeck1973}. From a theoretical point of view, the 2nd law
of Thermodynamics can be considered as fundamental as Mechanics, and
not as a mere phenomenological consequence of it as discussed e.g. by
\cite{doi:10.1142/S0218202515500566}, in the case of transport theory.

At the end of nineteenth century Boltzmann, although being conscious
about the autonomy of the 2nd law of Thermodynamics in respect to
Mechanics \citep[see][]{book:969643}, proposed to link the microscopic
properties of a gas, governed by Classical Mechanics, with its thermal
behavior expressed by the entropy. In other words, he connected, for
the first time, Classical Mechanics with Thermodynamics. This was done
through a kinetic equation, which refers to the distribution function
$f(\vec{x},\vec{v},t)$ in the general format
\begin{equation}
  \label{eq:1}
  \frac{df}{dt} = \Gamma[f],
\end{equation}
where the relaxation term $\Gamma[f]$ introduces the irreversibility
in the description. The so-called Boltzmann kinetic equation, which
applies to rarefied, short-range interacting (collisional) molecular
gases, can successfully describe the entropy increase and other
transport phenomena for these systems
\citep[see][]{cercignani1988boltzmann}.

After initially claiming to have derived the 2nd law of Thermodynamics
exclusively from Mechanics, Boltzmann had to recognize that his
equation contained extra, statistical content. This is not really an
artificial feature of his method, but rather a limitation of a purely
mechanical description of the evolution of large systems, which
cannot, \emph{per se}, make transparent the irreversible character of
the evolution of the macroscopic state expressed by the 2nd law of
Thermodynamics. In other words, an effective description of the
macroscopic evolution should take into account, besides Mechanics,
statistical ingredients to implement the time irreversibility in the
description \citep[see][]{1979wfst.book.....K}.

According to \cite{1993PhyA..194....1L}, arguments against the
physical reality of \emph{macroscopic} time irreversibility based on
the fact that individual particle trajectories are time reversible
were already satisfactorily answered in Boltzmann's times. A nice
review on this controversy, including references and an answer to the
issue, can be found in \cite{lebowitz2007time}. In fact, the simple
example of a gas expanding in a box is enough to illustrate the point:
suppose that all molecules of the gas are initially in a small region
around the center of the box with random velocities. Assume also that
one could turn off the collisions among the molecules, just allowing
for collisions with the walls (let us assume a spherical
container). The system would start evolving with each molecule in
uniform motion spreading in the box, then colliding with the walls and
occupying all the volume. Unless the system is prepared with very
specific initial conditions, it will never (or only after an extremely
long time) go back spontaneously to the initial state, even if all
trajectories are regular. There is macroscopic irreversibility and
entropy increase. However, there is no microscopic physical mechanism
randomizing the trajectories and driving the system towards a unique,
well-defined, equilibrium state. Instead, the system achieves an
invariant non-equilibrium state, which can keep some ``memory'' of the
initial state. This is a simple illustration of what is traditionally
called \emph{phase mixing}.

Now, consider a more realistic gas, in which the molecules are allowed
to collide amongst each other (in normal conditions, a typical
molecule can collide $\approx 10^6$ times per second!). The system
starts evolving with the molecules elastically colliding with each
other and with the walls. After a short time the entire box is filled
and the thermodynamical equilibrium is achieved, with the macroscopic
state being characterized by the Maxwell-Boltzmann distribution. The
system never (or extremely rarely) returns to the initial state with
all molecules in the center with the same initial velocities. In this
case also there is macroscopic irreversibility and entropy
increase. However, now the random collisions drive the system towards
a unique equilibrium state independent of the initial condition. This
is an extreme example of what is called \emph{chaotic mixing}, which
can occur even for collisionless evolution of ensembles in
non-integrable potentials \citep[see][]{Merritt_1996_2}.

Since the presence of chaos implies the existence of a predictability
horizon, i.e. of a time limit beyond which a trajectory cannot be
predicted with certainty, chaotic motion can be seen as introducing
irreversibility at a \emph{microscopic} level, because reversing the
velocities at any instant beyond the predictability horizon, is not
guaranteed to recover the initial state. The collisions amongst the
molecules in the example above represent an efficient (but non-unique)
mechanism to produce chaos in many-body systems.

In the last few decades, significant progress has been made in the
study of non-integrable systems \citep[see][]{Lichtenberg} and some
authors support the idea that chaos plays a fundamental role in the
macroscopic irreversible evolution. The most radical line of thought
in this direction is that of Prigogine \citep[see
e.g.][]{1999PhyA..263..528P}, who proposes that, in order to explain
the observed irreversibility at the macroscopic level, the time
irreversibility must be formally present on the microscopic level,
associated to a fundamental indeterminacy due to chaos, requiring a
deep revision of the fundamentals of Classical Mechanics.

On the other hand, according to \cite{1993PhyA..194....1L}, even
though ``instabilities induced by `locally' chaotic behavior do play a
role in determining the nature of the macroscopic evolution (...), the
central role in time asymmetric behavior is played by the very large
number of degrees of freedom involved in the evolution of macroscopic
systems''  \citep[see also][]{1999PhyA..263..516L}.

\section{Collisionless relaxation and chaotic vs regular phase mixing}
\label{sec:violent_relax}
Going back to the evolution of collisionless self-gravitating systems,
let us remember that their early and fast relaxation is
\emph{macroscopically} time irreversible. In order to avoid confusion
about this point, we emphasize that, even if we only consider the
$N$-body gravitational problem of dark matter halos in simulations,
neglecting any dissipative baryonic component, the process of galaxy
formation is macroscopically time irreversible because we only observe
this process occurring in one time direction. Besides that, the
gravitational $N$-body problem is intrinsically unstable
\citep[see][]{1964ApJ...140..250M, 2002ApJ...580..606H}, and the fast
relaxation of collapsing structures is expected to be accompanied by a
large amount of chaotic orbits
\citep[see][]{2003MNRAS.341..927K}. Interestingly,
\cite{2007ApJ...658..731V} have found that the presence of chaotic
orbits in simulated galaxy mergers seem to be associated mostly to the
Miller's instability than to a time-varying potential.

Note also that the violent relaxation scenario proposed by
\cite{LyndenBell_1967} is strongly based on the presence of a
time-varying potential. However, the need of such time dependence is
criticized by several authors \citep[see][]{Kandrup_1993, NYAS:NYAS28,
  2005NYASA1045....3M}. Additionally, \cite{doi:10.1093/mnras/stt935}
argues that it is possible to use a suitable coordinate frame in which
the potential remains ``static'', erasing any dynamical effect of this
time dependence.

Moreover, it is well-known that the final state of $N$-body
simulations depends on the initial state, only forming structures
resembling the observed ones when starting with cold (low velocity
dispersion) initial conditions \citep[see][]{1982MNRAS.201..939V,
  1984MNRAS.209...15M, 1984ApJ...281...13M}. On the other hand, some
seemingly universal properties do emerge from $N$-body simulations,
such as the NFW density profile \citep[see][]{NFW_1997, Navarro_2004}
also observed in real systems \citep[see][]{Umetsu_2011_b} or the
pseudo-phase-space density power-law
\citep[see][]{doi:10.1111/j.1365-2966.2011.19008.x}.

Following this discussion, it is possible to identify at least four
different ingredients that can contribute to the macroscopic time
irreversibility in the fast collisionless relaxation of $N$-body
gravitating systems (although not all of them need to operate
simultaneously):
\begin{enumerate}
\item{a large number of degrees of freedom,}
\item{phase mixing of particles with a spread of initial conditions,}
\item{the presence of chaotic mixing,}
\item{a time-dependent self-consistent potential.} 
\end{enumerate}

The orbit integration of ensembles in fixed external potentials
performed in this work allows us to investigate the possible roles of
ingredients $1-3$. Interestingly, the results shown in
\S~\ref{sec:plummer} and \S~\ref{sec:ellipsoid}, where we clearly
observe macroscopic irreversibility even in static and integrable
potentials, seem to be in line with the ideas of
\cite{1993PhyA..194....1L} quoted above. The important differences
introduced by the presence of chaotic orbits are discussed in
\S~\ref{sec:ellipsoid}.

\section{Entropy estimators}
\label{sec:estimators}
In \citetalias{2017ApJ...846..125B}, $N$-body simulations of
self-gravitating systems were run, starting with initial
configurations far from equilibrium and the evolution of the entropy
defined by Eq.~\eqref{eq:S_def} was studied. Following
well-established mathematical prescriptions, this entropy is estimated
as
\begin{equation}
  \label{eq:S_estimate_0}
  \hat{S}(t) = -\frac{1}{N}\sum_{i=1}^N \ln\hat{f}_i,
\end{equation}
where the integral over the phase-space is translated into a sum over
all the particles of the system. Of course, we still have the problem
of calculating $\hat{f}_i$, the estimate of the distribution function
$f$ at the position of each particle $i$. Eq. \eqref{eq:S_estimate_0}
has been shown to converge to Eq. \eqref{eq:S_def} for
$N\rightarrow \infty$ when we calculate $\hat{f}_i$ with at least two
methods \citep[see][for rigorous results]{Joe1989,Beirlant1997a,
  biau2015lectures}: the nearest neighbor and the kernel method. In
\citetalias{2017ApJ...846..125B} it is shown that both methods provide
very similar entropy evolutions, also similar to that obtained with
the EnBiD method (based on a phase-space tessellation into mutually
disjoint hypercubes) developed by \cite{2006MNRAS.373.1293S}.

In the nearest neighbor method, $\hat{f}_i$ is estimated as the number
of particles (one) inside a hyper-sphere of radius $D_{in}$ around the
particle $i$, divided by its volume. Including the normalization
factors \citep[see][]{Leonenko_2008}, we have generically in $d$
dimensions:
\begin{equation}
  \label{eq:f_NN}
  \hat{f}_i = \frac{1}{(N-1)e^\gamma V_d D_{in}^d},
\end{equation}
where $\gamma \approx 0.57722$ is the Euler-Mascheroni constant,
$V_d = \pi^{d/2}/\Gamma(d/2 + 1)$ is the volume of a hyper-sphere of
unitary radius and
\begin{equation}
  \label{eq:dist_phase_space}
  D_{in} = \sqrt{(\vec{r}_i - \vec{r}_n)^2 + (\vec{v}_i - \vec{v}_n)^2}
\end{equation}
is the distance in phase-space of particle $i$ to its nearest neighbor
$n$. Thus, in 6 dimensions we have
\begin{equation}
  \label{eq:S_estimate}
  \hat{S} = \frac{1}{N}\sum_{i=1}^N \ln D_{in}^6 +
  \ln\left[\frac{\pi^3}{6}(N-1) \right] + \gamma.
\end{equation}
In Eq.\eqref{eq:dist_phase_space}, it is assumed that positions and
velocities are represented with coordinates that are dimensionless and
have similar variances in different directions -- see
\citetalias{2017ApJ...846..125B} for more details. In this work, each
coordinate is normalized by its initial inter-percentile range
containing $68\%$ of the data around the median.

By means of a tree algorithm, it is possible to optimize the ``naive''
search for the nearest neighbor, which originally has complexity
$N^2$, decreasing the complexity to $N\ln N$ -- see
\cite{Friedman:1977:AFB:355744.355745}. In this work we restrict
ourselves to this method due to its speed and illustrative
simplicity. For the identification of the neighbors we use the kd-tree
algorithm \emph{Approximate Nearest Neighbor} (ANN) developed by
\cite{Arya:1998:OAA:293347.293348}\footnote{Available at
  {www.cs.umd.edu/$\sim$mount/ANN/}. A slightly different version,
  allowing searches in parallel, was developed by Andreas Girgensohn
  and kindly provided by David Mount.}. The algorithm allows to
optimize the search by approximating the nearest neighbor, but we use
it without any approximation, identifying the exact nearest neighbor.

In simple cases, it is possible to obtain analytic expressions for the
entropy defined by Eq.~\eqref{eq:S_def}, and to compare them with what
we get with the estimator Eq. \eqref{eq:S_estimate}. For example, for
an ensemble uniformly sampling a sphere of radius $r_{max}$ in
positions and $v_{max}$ in velocities, which is used as initial
condition in \S~\ref{sec:ellipsoid}, the distribution function is
independent of the coordinates and the analytic expression for the
entropy is
\begin{multline}
  \label{eq:S_uniform_sph}
  S_0 = -\int f_0\ln f_0 \,d^3\vec{r}\,d^3\vec{v} =\\
  = -\ln f_0 = \ln\left[ \left(
      \frac{4\pi}{3}\right)^2r_{max}^3v_{max}^3\right].
\end{multline}
For this configuration, with $N=10^6$ the estimator
Eq. \eqref{eq:S_estimate} provides an error of $\approx 4\%$ relative
to Eq. \eqref{eq:S_uniform_sph}.

Another simple case is for a distribution function depending on energy
only. In this case, Eq.~\eqref{eq:S_def} reduces to
\begin{equation}
  \label{eq:S_f_E}
  S = -\int_{\phi(0)}^0 f(E)g(E)\ln f \,dE,
\end{equation}
where $E$ is the energy per unit mass, $\phi(r)$ is the gravitational
potential and
\begin{equation}
  \label{eq:3}
  g(E) = 16\pi^2 \int_0^{r_m(E)}dr r^2 \sqrt{2(E - \phi(r))}
\end{equation}
is the density of states.  A simple example for which $f = f(E)$ is
the Plummer model, characterized by
Eqs.\eqref{eq:phi_plummer}-\eqref{eq:f_plummer} below. With $N=10^6$
particles sampling this model, the error provided by
Eq. \eqref{eq:S_estimate} relative to Eq. \eqref{eq:S_f_E} is
$\approx 0.25\%$. Appendix \S~\ref{sec:errors} discusses the
$N$-dependence of the estimator uncertainties.

\section{Harmonic potential and macroscopic reversibility}
\label{sec:harmonic}
We first study the dynamical evolution of gravitational systems
integrating orbits in the harmonic oscillator potential. This allows
us to verify if the entropy increase observed in
\citetalias{2017ApJ...846..125B} (and in the results shown below) can
be due to spurious truncation errors that could give rise to
macroscopic effects and artificially introduce time
irreversibility. The harmonic potential is given by
\begin{equation}
  \label{eq:pot_harm}
  \phi(r) = \frac{1}{2}\Omega^2 r^2,
\end{equation}
where $\Omega$ is the angular frequency and $r$ is the distance to the
center. In this special potential the angular (azimuthal) period,
$T = 2\pi/\Omega$, is the same for all particles, independently of
their energies. Thus, even though there is phase mixing within one
period, after one period each particle is back to its initial position
and velocity. Consequently, in addition to the usual
\emph{microscopic}, we also have \emph{macroscopic} time
reversibility.

We start our numerical simulations with $N=10^6$ particles sampling a
Plummer model, characterized by
Eqs.~\eqref{eq:phi_plummer}-\eqref{eq:f_plummer}. The sampling and
orbit integration in this potential, as well as the others discussed
in the following sections, were performed with the Agama Library
\citep[][]{2018MNRAS.tmp.2556V}. In this section, we integrate orbits
in the potential given by Eq.\eqref{eq:pot_harm} for 30 orbital
periods, setting $GM=1$ and $a=1$ in
Eqs. ~\eqref{eq:phi_plummer}-\eqref{eq:f_plummer}. The entropy is then
estimated at each snapshot with
Eqs.~\eqref{eq:f_NN}-\eqref{eq:S_estimate}. We repeated this procedure
for 10 different realizations and calculated the average of the
entropy at each snapshot.

\begin{figure}
  \raggedright
  \includegraphics[width=9.5cm]{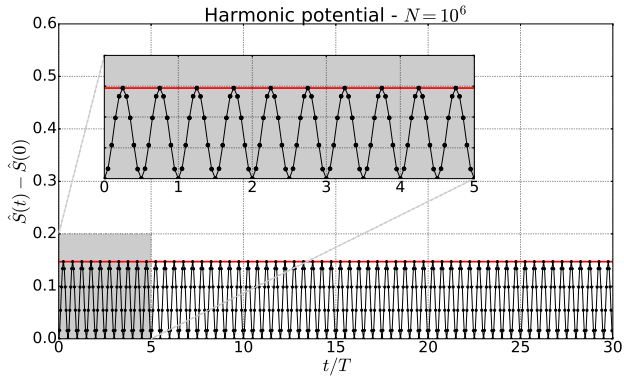}
  \vspace{-0.2cm}
  \caption{Entropy evolution obtained integrating $N=10^6$ orbits in
    the Harmonic potential, Eq. \eqref{eq:pot_harm}, which, besides
    microscopic time reversibility, also generates macroscopic
    reversibility. The horizontal red line shows that there is no net
    entropy increase and the inset upper plot shows the oscillatory
    pattern, making clear that the macroscopic time reversibility is
    perfectly recovered by the numerical scheme.}
  \label{fig:S_harm}
\end{figure}

The result is shown in Fig.\ref{fig:S_harm}. The uncertainties,
estimated as the mean standard deviation over 10 realizations, are
$\sigma_{\Delta \hat{S}}\approx 0.001$ (smaller than the data
points). The entropy keeps oscillating with a constant maximum
amplitude (red horizontal line), without any global increase (the
difference in amplitude between the last and first peaks is
$\approx 10^{-7}$). Zooming-in (upper inset plot) helps to visualize
the oscillatory pattern. Note that, due to the spherical symmetry, the
system returns to the initial macroscopic state after one
\emph{radial} period (when the particles are in positions
diametrically opposite to the initial ones), which is half the angular
period $T$, and the entropy completes two cycles in each period.

Thus we conclude that for systems whose evolution is
\emph{macroscopically} time reversible (a highly exceptional
situation), our numerical procedure is able to perfectly recover this
reversibility. This shows that there is no information loss in the
orbit integration or in the entropy estimation that could give rise to
an artificial entropy production. This simple example also makes
explicit the difference between microscopic time reversibility, which
is always theoretically present, at the level of single trajectories,
and macroscopic time reversibility, which is present in this very
particular example but not in general systems such as the ones
discussed below.

\section{Plummer potential and macroscopic irreversibility}
\label{sec:plummer}
We now use the Plummer potential to integrate two different initial
conditions. The first is generated with $N=10^6$ particles sampling a
uniform sphere of radius $a$ and maximum velocity
$v_{max} = \sqrt{2|\phi(a)|}$, where $a$ is the Plummer scale radius
of a system of mass $M$. In this case, $v_{max} = \sqrt{\sqrt{2}GM/a}$
and
\begin{equation}
  \label{eq:phi_plummer}
  \phi(r) = -\frac{GM}{a}\frac{1}{\sqrt{1 + (r/a)^2}}.
\end{equation}

We then integrate these orbits in the potential given by
Eq.\eqref{eq:phi_plummer} for 30 crossing times, setting $a=1$ and
$GM = 1$. The crossing time was estimated as
\begin{equation}
\label{eq:tau_cr}
\tau_{cr} = 2\pi \sqrt{\frac{\langle r^2\rangle}{\langle v^2\rangle}},
\end{equation}
where the averages are calculated at $t=0$. The entropy is again
estimated at each snapshot with
Eqs.~\eqref{eq:f_NN}-\eqref{eq:S_estimate}, repeating the procedure
for 10 different realizations and calculating the entropy average at
each snapshot. In Appendix \ref{sec:errors} we show the $N$-dependence
on both systematic (bias) and statistical errors (normal fluctuations)
in the entropy estimators for these and other initial conditions.

\begin{figure}
  \raggedright
  \includegraphics[width=9.5cm]{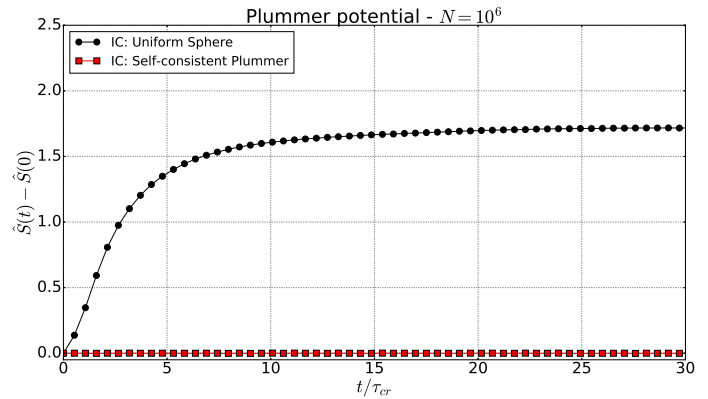}
  \vspace{-0.2cm}
  \caption{Entropy evolution obtained integrating orbits in the
    Plummer potential, Eq.~\eqref{eq:phi_plummer}. Black points:
    initial condition (IC) sampling a uniform sphere. Red squares: IC
    sampling a stationary state of the Plummer model,
    Eq.~\eqref{eq:f_plummer}. As expected from the 2nd law of
    Thermodynamics, when starting far from equilibrium, the entropy
    increases until it achieves a maximum determined by the
    constraints. The entropy increases not due to numerical errors
    since it is correctly conserved when the initial state is
    stationary.}
  \label{fig:S_plummer}
\end{figure}

The resulting entropy evolution is shown in Fig.\ref{fig:S_plummer}
(black dots). Since the initial condition is far from an invariant
state and because the Plummer potential does not share with the
Harmonic potential the very particular property of generating
\emph{macroscopic} reversibility, the system evolves through phase
mixing and the entropy increases until the system achieves an
invariant state, just as expected from the 2nd law of Thermodynamics.

The fact that the entropy increases in a few crossing times is similar
to what is observed in \citetalias{2017ApJ...846..125B} for the
evolution of self-gravitating $N$-body simulations. While in the
latter case, given the large numbers of particles, two-body relaxation
was expected to be negligible for this short timescale, in the present
work there is by definition no two-body relaxation, since we integrate
independent orbits in an external potential. Note that the entropy
increase occurs despite the Plummer potential being time independent
and integrable (like any spherically symmetric model), i.e. despite
the absence of chaotic orbits. The role of chaos for the time
irreversibility is explored in \S~\ref{sec:ellipsoid}.

The second initial condition was generated by sampling a Plummer model
of same mass and scale radius as in the Plummer potential used for
integration, for which the stationary state is given by
\citep[see][]{1974A&A....37..183A}:

\begin{equation}
  \label{eq:f_plummer}
  f(E) = \frac{24\sqrt{2}}{7\pi^3}\frac{a^2}{(GM)^5}(-E)^{7/2}.
\end{equation}
Since this initial condition represents by definition a stationary
state of the potential used for integration, the entropy should be
conserved during the evolution of the system. Indeed,
Fig.\ref{fig:S_plummer} shows that the entropy (red squares) is
perfectly conserved
${(|\hat{S}(30\tau_{cr}) - \hat{S}(0)|/\hat{S}(0) \approx0.015\%)}$.
This shows again that the entropy increase observed in other
configurations is not introduced by information loss due to truncation
errors or to errors in the entropy estimation, otherwise an artificial
entropy production would be very likely observed even for a stationary
state.

\section{Ellipsoidal model and $N$-dependence}
\label{sec:ellipsoid}
Having shown for integrable systems that the entropy estimates have
qualitative agreement with the 2nd law of Thermodynamics and
quantitative agreement for initial states where the distribution
function is known, we now investigate the entropy evolution in
\textit{non-integrable} ellipsoidal potentials. The $N$-dependence of
these estimates sheds light on the role of chaos for time
irreversibility.

The model used is defined by the density profile
\begin{equation}
  \label{eq:rho_ellipsoid}
  \rho(m) = \rho_0m^{-\gamma}\left (1 + m^2\right)^\frac{\gamma - 4}{2},
\end{equation}
where $m^2 = x^2/a^2 + y^2/b^2 + z^2/c^2$, $a$, $b$ and $c$ being the
semi-axis in the respective directions, and $\rho_0$ is a normalizing
factor. We fix the total mass to be $M=1$ and scale $a=1$,
$b/a = 0.8$, $c/a = 0.5$.  All models considered here are thus
triaxial, with a density profile $\rho(m)\propto m^{-4}$ in the
external regions. Moreover, when $\gamma = 0$, this reduces to the
so-called Perfect Ellipsoid, which is fully integrable (i.e. all
orbits are regular) despite being triaxial -- see
\cite{1985MNRAS.216..273D}. On the other hand, the introduction of a
cusp with inner slope $\gamma > 0$ breaks integrability, giving rise
to chaotic orbits. \cite{1998ApJ...506..686V}, in a fundamental
frequency analysis of orbits evolved in a triaxial
\cite{1993MNRAS.265..250D} model, concluded that the fraction of
chaotic orbits increases with the central slope $\gamma$ and that a
transition to global stochasticity occurs for ${\gamma \gtrsim
  2}$.
The analysis performed in \S\ref{sec:freq_analysis} drives us to
similar conclusions for the models used here.

We evolve orbits in this model with $\gamma = 0$ (Perfect Ellipsoid),
$\gamma = 1$ (weak cusp) and $\gamma = 2$ (strong cusp). These values
cover the inner slopes observed in galaxies, dark matter halos and
galaxy clusters. We use two different initial conditions: first, a
sphere of radius $a$ and maximum velocity
${v_{max}=\sqrt{2|\phi(0,0,a)|}}$ uniformly sampled in positions and
velocities; second, a Gaussian distribution with standard deviation in
the spatial coordinates $\sigma_r = a/3$ and an independent Gaussian
distribution for the velocity components with $\sigma_v = v_{max}/3$,
both truncated at $3\sigma$, (i.e. at $a$ and $v_{max}$ for spatial
and velocity coordinates respectively).

\subsection{The structure of phase-space}
\label{sec:freq_analysis}

Before investigating the entropy evolution, we study the phase-space
structure associated with the models given by
Eq.~\eqref{eq:rho_ellipsoid}. By means of a frequency analysis (in
Cartesian coordinates), we identify the fundamental frequencies,
resonances and fractions of regular and chaotic orbits
\citep[see][]{1998ApJ...506..686V, 1999PASP..111..129M,
  2016MNRAS.455.1079P}. This can be seen as the microscopic
counterpart of the macroscopic characterization made with the entropy
estimates. Note, however, that there is no reason to consider the
former as more fundamental than the latter. Instead, these can be seen
as complementary approaches.

An orbit evolved in an Hamiltonian $H$ with $N$ degrees of freedom can
be described with the time evolution of $2N$ components
$(x_1(t),...,x_{2N}(t)) = (q_1(t),...,q_{N}(t),p_1(t),...,p_{N}(t))$,
where
\begin{equation}
  \label{eq:hamilton}
  \dot{q}_k = \frac{\partial H}{\partial p_k}, \qquad
  \dot{p}_k = -\frac{\partial H}{\partial q_k}.
\end{equation}
Each component of a bounded orbit can be written as
\begin{equation}
  \label{eq:series_quasi_period}
  x(t) = \sum_{j=1}^\infty A_j e^{i\omega_j t},
\end{equation}
where $A_j$ are complex amplitudes. If there are at least $N$
isolating integrals of motion, every orbit is regular and restricted
to a surface of dimension $\leq N$. In this case, it is possible to
apply a global canonical transformation to define angle-action
coordinates $(\vec{\theta}, \vec{J})$ such that the Hamiltonian
resembles that of free particles, i.e. it only depends on the momenta
$\vec{J}$: $H = H(\vec{J})$. In this way, Eqs.~\eqref{eq:hamilton}
imply that $\vec{J}$ is constant and
$\vec{\theta}\propto \vec{\Omega}t$, with constant fundamental
frequencies $\vec{\Omega} = (\Omega_1,...,\Omega_N)$.  Moreover, the
frequencies $\omega_j$ in Eq.~\eqref{eq:series_quasi_period} can be
written as a linear combination of the fundamental frequencies, i.e
${\omega_j = \vec{n}\cdot\vec{\Omega}}$, where $\vec{n}$ is a vector
with $N$ integer components (we will assume a system with $N=3$
degrees of freedom, the $3$ spatial coordinates of each orbit).

Computing the Fourier transform of each of the components $x(t)$
integrated over $\approx 100 \tau_{cr}$, it is possible to identify
discrete peaks whose locations, combined with the amplitudes $A_j$,
can be used to obtain the leading frequencies $\vec{\Omega}$
\citep[][]{1982ApJ...252..308B}. A fast and accurate technique to
calculate the three leading frequencies in any coordinate system was
developed by \cite{1990Icar...88..266L}, and an implementation that
uses integer programming \citep{1998ApJ...506..686V} is used in this
work.

If the coordinate system used to compute the lead frequencies is close
to the angle variables associated with an orbit, then the lead
frequencies obtained by the above method are the fundamental
frequencies (i.e. time derivatives of the angle variables). A previous
work \citep{1998ApJ...506..686V} has shown that for box orbits, in a
Cartesian system with the $x,y,z$ axes aligned with the long,
intermediate and short axes of the triaxial ellipsoid, the coordinates
are adequately close to the angle variables, and thus
$\Omega_x, \Omega_y, \Omega_z$ are the fundamental frequencies for box
orbits. However for short and long axis tube orbits it is necessary to
compute orbital frequencies in symplectic polar coordinates with the
symmetry axis of the coordinate system aligned with the symmetry axis
of the tube ($z$ axis for short-axis tubes and $x$ axis for long-axis
tubes). If the frequencies of tube orbits are computed in Cartesian
coordinates, they appear as lines in frequency maps, although they are
not resonances (for more on resonances see below). Therefore, we limit
our discussion of frequency maps to box orbits.

\begin{figure*}
  \epsscale{0.85}
  \plotone{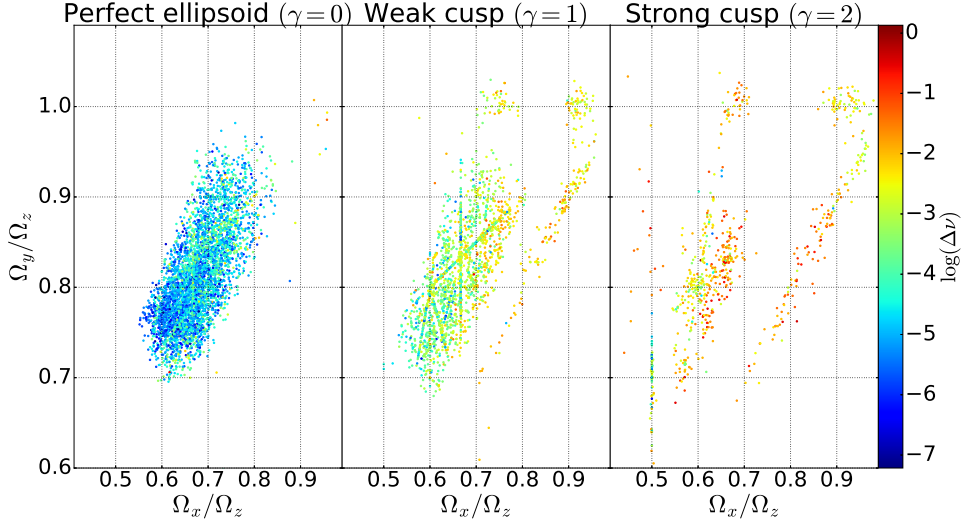}
  \caption{Frequency map of the box orbits selected from $10^4$ orbits
    with the uniform sphere initial conditions, for the three models
    $\gamma=0,1,2$, color coded by $\log \Delta \nu$. The Perfect
    Ellipsoid only generates regular orbits (small $\Delta \nu$),
    while larger inner slopes $\gamma$ generates increasing fractions
    of chaotic orbits. Note also the prominence of resonant lines in
    the weak cusp model.}
  \label{fig:freq_map_10k_unif_sph}
\end{figure*}

\begin{figure*}
  \epsscale{0.85}
  \plotone{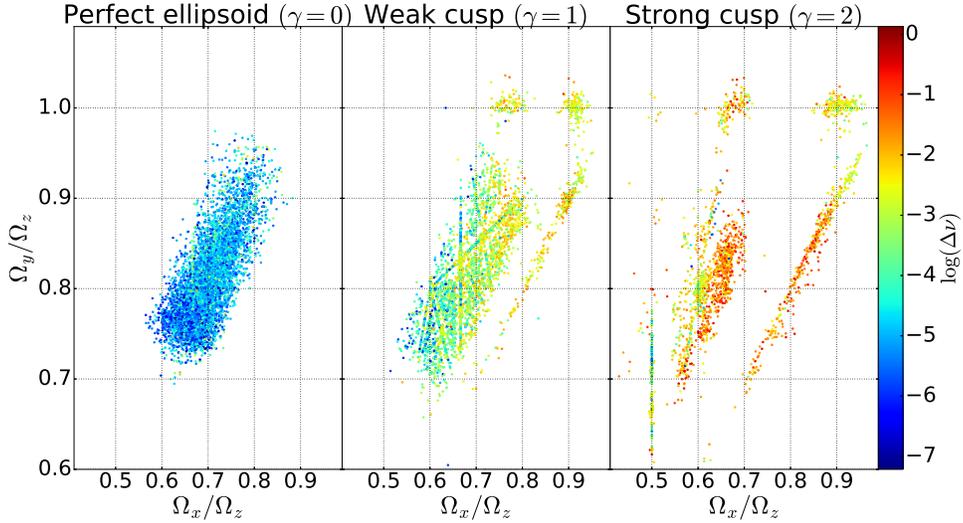}
  \caption{Similar to Fig.~\ref{fig:freq_map_10k_unif_sph}, but now
    for the Gaussian initial conditions. Once more, the Perfect
    Ellipsoid only generates regular orbits, while larger inner slopes
    $\gamma$ generate increasing fractions of chaotic orbits. Note
    again the prominence of resonant lines in the weak cusp model.}
  \label{fig:freq_map_10k_gauss}
\end{figure*}

A frequency map is a plot showing ratios of these frequencies for each
orbit -- see
Figs.~\ref{fig:freq_map_10k_unif_sph}-\ref{fig:freq_map_10k_gauss}. In
general, the fundamental frequencies are independent. However, for
some orbits (resonances), the fundamental frequencies can be such that
$\vec{n}\cdot\vec{\Omega} = 0$. When two such conditions are
satisfied, the orbit is closed (periodic). In a frequency map, stable
resonances appear as filled straight lines, while unstable resonances
appear as void lines.

In an integrable Hamiltonian $H^0$, the fundamental frequencies of
each orbit are uniquely determined by its initial conditions and are
conserved. In this case, in general initial conditions, resonances
appear only ``by chance'', representing a set of measure zero. In a
non-integrable model generated by a small perturbation of $H^0$, it is
common to observe the phenomenon of resonance trapping, in which
orbits close to a resonant condition get ``captured'' by the resonance
\citep[][]{Binney_2008}. A consequence of resonance trapping is a
further restriction of the phase-space region explored by the
orbit. The relation of this effect to relaxation is discussed in \S
\ref{sec:entropy_gauss_ic}.

A related important aspect of resonances for the phase-space structure
can be illustrated by the Kolmogorov-Arnold-Moser (KAM) theorem
\citep[see][]{Lichtenberg}. According to it, if an integrable
Hamiltonian is perturbed, the very non-resonant orbits maintain their
topological properties, i.e. remain quasi-periodic. The same happens
for orbits close to stable resonances. On the other hand, orbits close
enough to unstable resonances can be drastically modified, giving rise
to stochastic motion even for a small perturbation (thus, resonances
can be seen as seeds of stochasticity). With a small perturbation,
stochastic regions are separated from each other, and when the
perturbation is increased these regions tend to grow and to overlap,
eventually transitioning to global stochasticity
\citep[see][]{cohen1975fundamental, CHIRIKOV1979263, Lichtenberg}.

While for regular orbits the fundamental frequencies identified by the
algorithm are conserved, this is not the case for chaotic
orbits.\footnote{Although the leading frequencies in Cartesian
  coordinates are not fundamental frequencies for tube orbits, they
  are still conserved since the true fundamental frequencies are
  linear combinations of the lead frequencies. Hence in what follows
  we focus on fundamental frequencies.} In order to classify the
orbits, this fact can be explored by computing the fundamental
frequencies in two consecutive time intervals $T_1$ and $T_2$. For
each orbit we compute
\begin{equation}
  \label{eq:5}
  \Delta \nu_i = \frac{\Omega_i(T_2) - \Omega_i(T_1)}{\Omega_i(T_1)},
\end{equation}
where $i$ refers to each component. Then we define the ``frequency
drift'' $\log (\Delta \nu)$ as the largest among the three. The larger
$\log (\Delta \nu)$, the more chaotic the orbit.

We evolve $N=10^4$ orbits for $\approx 100\tau_{cr}$, recording their
phase space coordinates at $10^5$ equally spaced time
steps\footnote{\cite{1998ApJ...506..686V} showed that increasing the
  integration time and decreasing the time spacing in the time series
  increases the accuracy for the recovered orbital frequencies in an
  integrable potential; for our current orbit integrations,
  $\log(\Delta \nu) < -4$ for the majority of orbits in the Perfect
  Ellipsoid which should contain only regular orbits. For a fraction
  of orbits with longer orbital periods than the average $\tau_{cr}$
  and very small fundamental frequencies the accuracy with which the
  numerical scheme recovers the frequencies is lower. This can be the
  reason for the appearance of hints of resonances in the left panels
  of Figs.~\ref{fig:freq_map_10k_unif_sph} and
  \ref{fig:freq_map_10k_gauss}.}. Then, the box orbits are
selected. Fig. \ref{fig:freq_map_10k_unif_sph} shows the frequency map
obtained with the uniform sphere initial conditions in the three
models $\gamma=0,1,2$, color coded by the value of $\log \Delta \nu$.
As expected, the Perfect Ellipsoid essentially generates only regular
orbits, showing that the numeric scheme for orbit integration and
frequency identification is accurate. Larger inner slopes $\gamma$
produce increasing fractions of chaotic orbits, in agreement with the
conclusions drawn by \cite{1998ApJ...506..686V} for a triaxial Dehnen
model.

The presence of several resonantly trapped orbits is clearly evident
as straight lines in the frequency maps of the two non-integrable
models $\gamma = 1,2$ (recall that resonant orbits satisfy a condition
like $l\Omega_x+m\Omega_y+n\Omega_z =0$, where $l$, $m$, $n$ are
integer numbers). In the weak cusp model ($\gamma = 1$) we see
numerous stable resonances which appear as clusters of points along
straight lines. It is clear that they are stable because their
diffusion rates (as indicated by the colors of the points) are small
(typically $\log(\Delta \nu) \lesssim -3$). The orange points
scattered along a line with a slope of approximately unity on the
right-hand side of the maps correspond to chaotic box orbits which are
associated with the stochastic layer (separatrix) between the family
of short-axis tubes with $\Omega_x \sim \Omega_y$.  Similarly the
clusters of orange points at $\Omega_y/\Omega_z \sim 1$ and
$\Omega_x/\Omega_z \sim 0.75$ arise from the chaotic box orbits
associated with the stochastic layer between the inner and outer
long-axis tubes, while those at $\Omega_y/\Omega_z \sim 1$ and
$\Omega_x/\Omega_z \sim 0.9$ are associated with the stochastic layer
between the outer long-axis tubes and the short-axis tubes.

Previous studies have shown that when an integrable potential is
perturbed by a central density cusp \citep[as in this paper
and][]{1998ApJ...506..686V}, or a central supermassive black hole or
figure rotation \citep{deibel_etal_11}, both the strength and number
of resonances increase. As the strength of the perturbation increases
(e.g. from $\gamma=1$ to $\gamma=2$), the resonances grow stronger and
begin to overlap. This is accompanied by an overlap of the
separatrices surrounding the resonances (which contain hyperbolically
unstable orbits). Resonance overlap is a well known cause of global
chaos in Hamiltonian systems \citep{CHIRIKOV1979263} and may be
thought of as occurring when several different resonances compete to
trap the same orbit \citep{Binney_2008}. This accounts for the fact
that the frequency map for the $\gamma=2$ model appears to have very
few regular regions and is largely occupied by chaotic orbits.

Fig.~\ref{fig:freq_map_10k_gauss} shows the frequency map for the box
orbits selected from the sample with the Gaussian initial conditions,
again color coded by $\log (\Delta \nu)$. The qualitative behavior is
very similar to the previous case: in the Perfect Ellipsoid all the
orbits can be safely classified as regular, while larger inner slopes
$\gamma$ generates increasing fractions of chaotic orbits. As in the
previous case, the weak cusp model $\gamma=1$ shows a prominence of
resonance lines, which are destroyed in the strong cusp model
$\gamma = 2$.

\begin{figure}
  \raggedright
  \includegraphics[width=9.5cm]{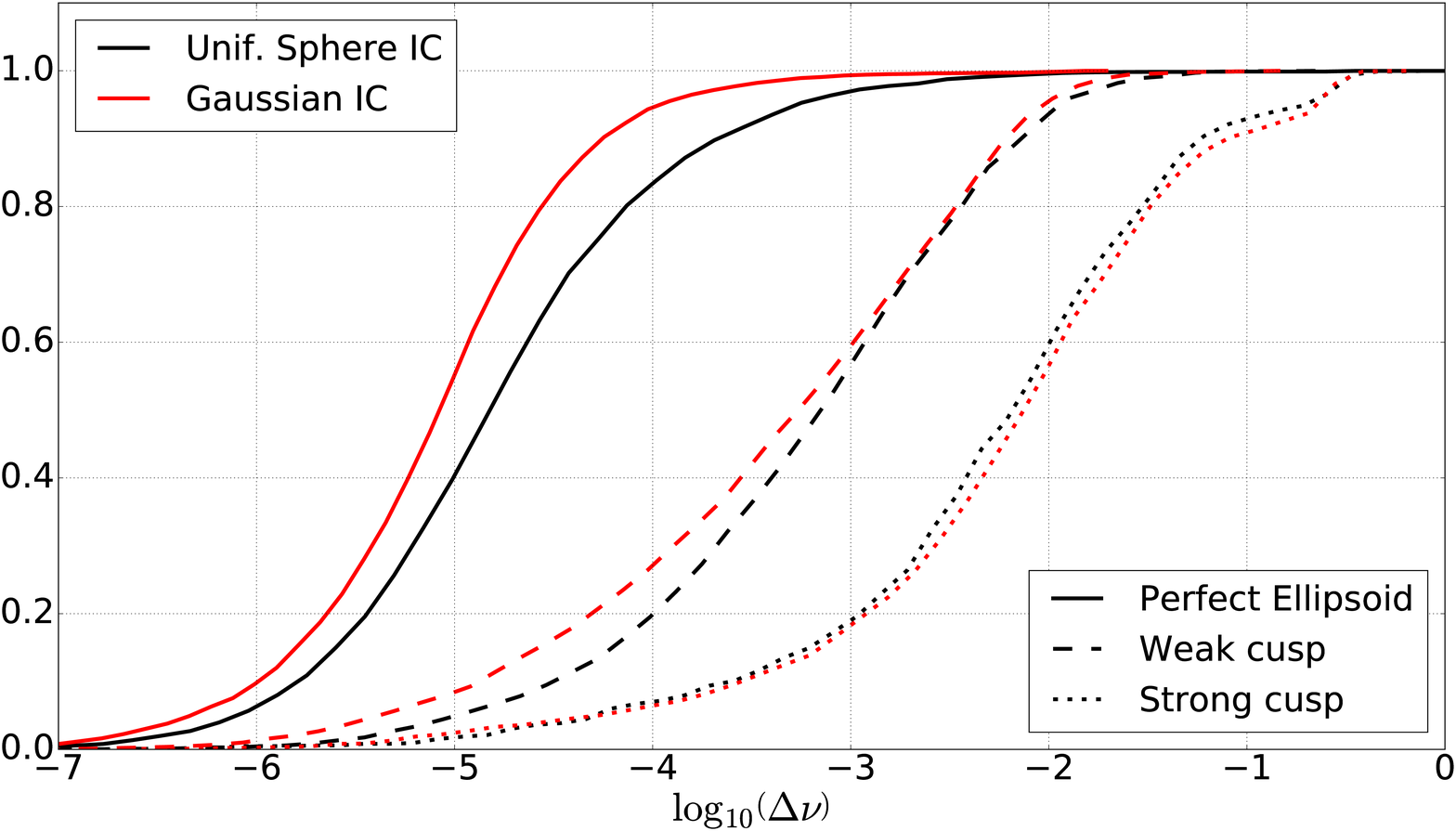}
  \vspace{-0.2cm}
  \caption{Cumulative distribution function for the frequency drift
    $\log \Delta \nu$. Larger inner slopes $\gamma$ produces increasing
    fractions of chaotic orbits.}
  \label{fig:cdf_log_f}
\end{figure}

In Fig.~\ref{fig:cdf_log_f} we show the cumulative distribution
function for the frequency drift $\log (\Delta \nu)$ in the three models
($\gamma=0,1,2$) and for the two initial conditions (different
colors). This plot summarizes the main conclusion from the previous
results: the introduction of larger inner slopes produces increasing
fractions of chaotic orbits. Additionally,
Figs.~\ref{fig:freq_map_10k_unif_sph}-\ref{fig:freq_map_10k_gauss}
show that a large fraction of orbits get trapped in resonances when
evolved in the weak cusp model $\gamma=1$. Having established these
results, we now study the entropy evolution in these models,
integrating the two initial conditions for $300\tau_{cr}$, with
numbers of orbits ranging from $N=10^4$ to $N=10^8$.

\subsection{Entropy evolution: uniform initial conditions}
\begin{figure*}
  \epsscale{0.85}
  \plotone{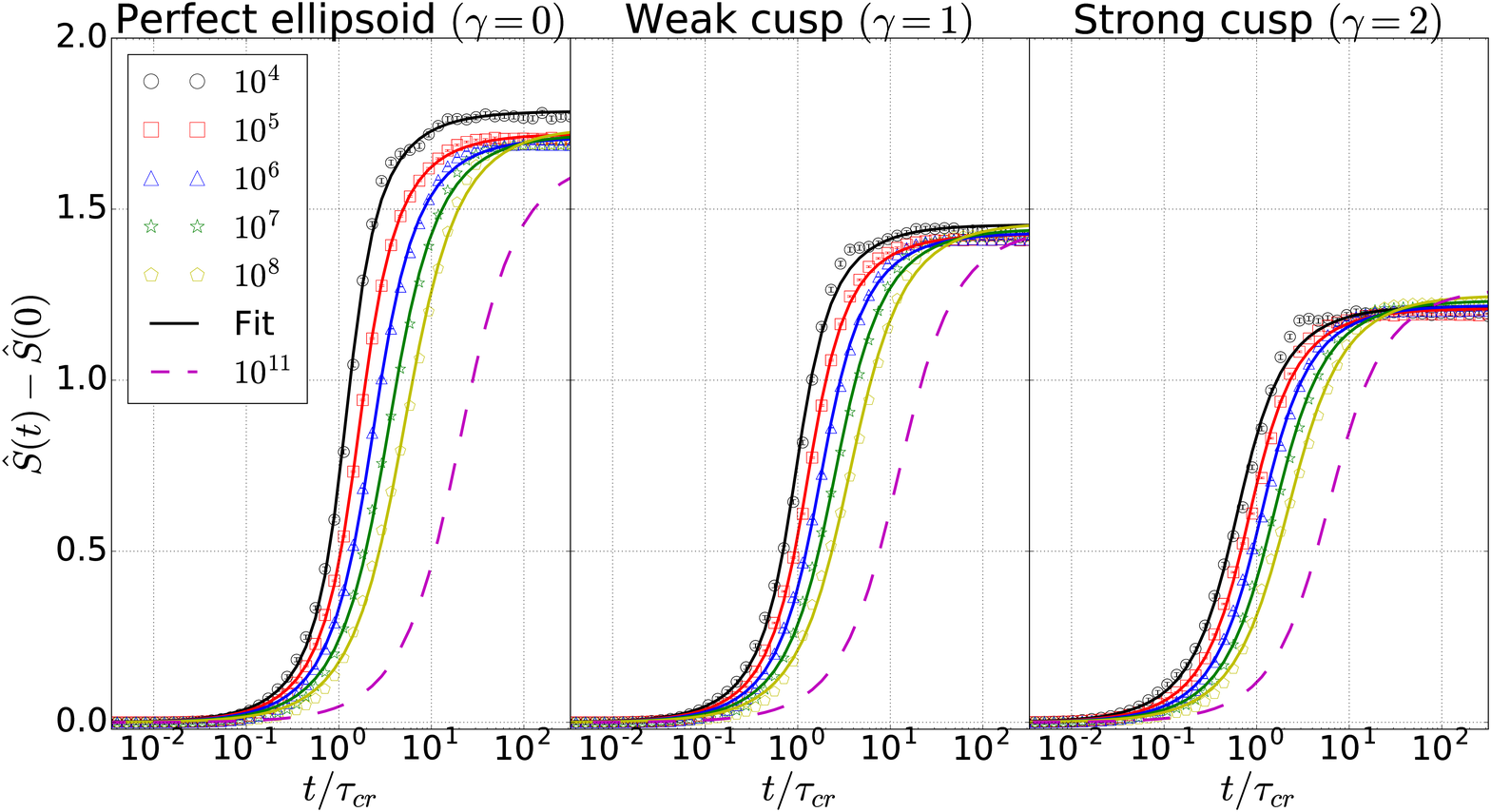}
  \caption{Entropy evolution for the initial uniform sphere evolved in
    the triaxial model defined by Eq.~\eqref{eq:rho_ellipsoid}, for
    inner density slopes $\gamma=0,1,2$. For all models the entropy
    achieves a maximum in $\approx 10-50\tau_{cr}$. Solid lines show
    best fits of Eq. \eqref{eq:fit_delta_S}, and dashed lines
    represent the extrapolation to ${N=10^{11}}$, using the parameters
    power laws $N$-dependences obtained in
    Fig.~\ref{fig:params_unif_sph_ellips}.}
  \label{fig:S_unif_sph_ellips}
\end{figure*}

The data points in Fig. \ref{fig:S_unif_sph_ellips} show the entropy
evolution for orbits integrated in the three potential models
($\gamma=0,1,2$) with the uniform sphere initial condition. Different
colors represent ensembles with different numbers of orbits. The
points represent an average over 10 realizations for $N\leq 10^6$, but
only one for $N\geq 10^7$.

In all models the entropy increases rapidly, achieving a maximum after
$\approx 10-50\tau_{cr}$. The maximum entropy value is different for
each model, which is not surprising since the phase-space volume
accessible to each ensemble depends on the model and the nature of the
orbits comprising the ensemble. In the integrable potential
($\gamma = 0$), each orbit explores the entire surface of a torus in
phase-space. For $\gamma = 1, 2$ we have significant fractions of
chaotic orbits and since such orbits (in a Hamiltonian potential) only
conserve one integral of motion (energy) they will undergo rapid
chaotic mixing (in $\sim 30-100\tau_{cr}$ as pointed out by
\cite{Merritt_1996_2}) to fill the 5-dimensional phase space surface
defined by the energy. In practice in most non-integrable potentials,
orbits remain trapped in lower dimensional regions of phase-space
defined by resonances (the so called Arnold web) that do not
correspond to a unique state associated with thermodynamical
equilibrium, or even to a stationary state self-consistently related
to the potential.

The data points in Fig.~\ref{fig:S_unif_sph_ellips} are fitted by the
function
\begin{multline}
  \label{eq:fit_delta_S}
  \Delta \hat{S} = \frac{A}{\pi/2 + \arctan
    \left(BC\right)}\times\\
  \times \left\{\arctan\left[B\left(t/\tau_{cr} - C\right)\right] + \arctan
    \left(BC\right)\right\},
\end{multline}
where parameters $A$, $B$ and $C$ represent respectively: the final
entropy increase, the relaxation rate and the time delay (in units of
$\tau _{cr}$) for entropy production, i.e. the time at which the
entropy starts to increase. The terms $\arctan \left(BC\right)$ and
$\pi/2 + \arctan \left(BC\right)$ ensure that
${\Delta \hat{S} (t=0)=0}$ and that
$A = \Delta \hat{S} (t\rightarrow \infty)$, respectively. Note that
$C$ gives an upper bound on the timescale for which the entropy
evolution is approximately compatible with the Vlasov equation, which
yields no entropy production. This point is discussed in
\S\ref{sec:meaning_vlasov}.

\begin{figure*}
  \epsscale{0.85}
  \plotone{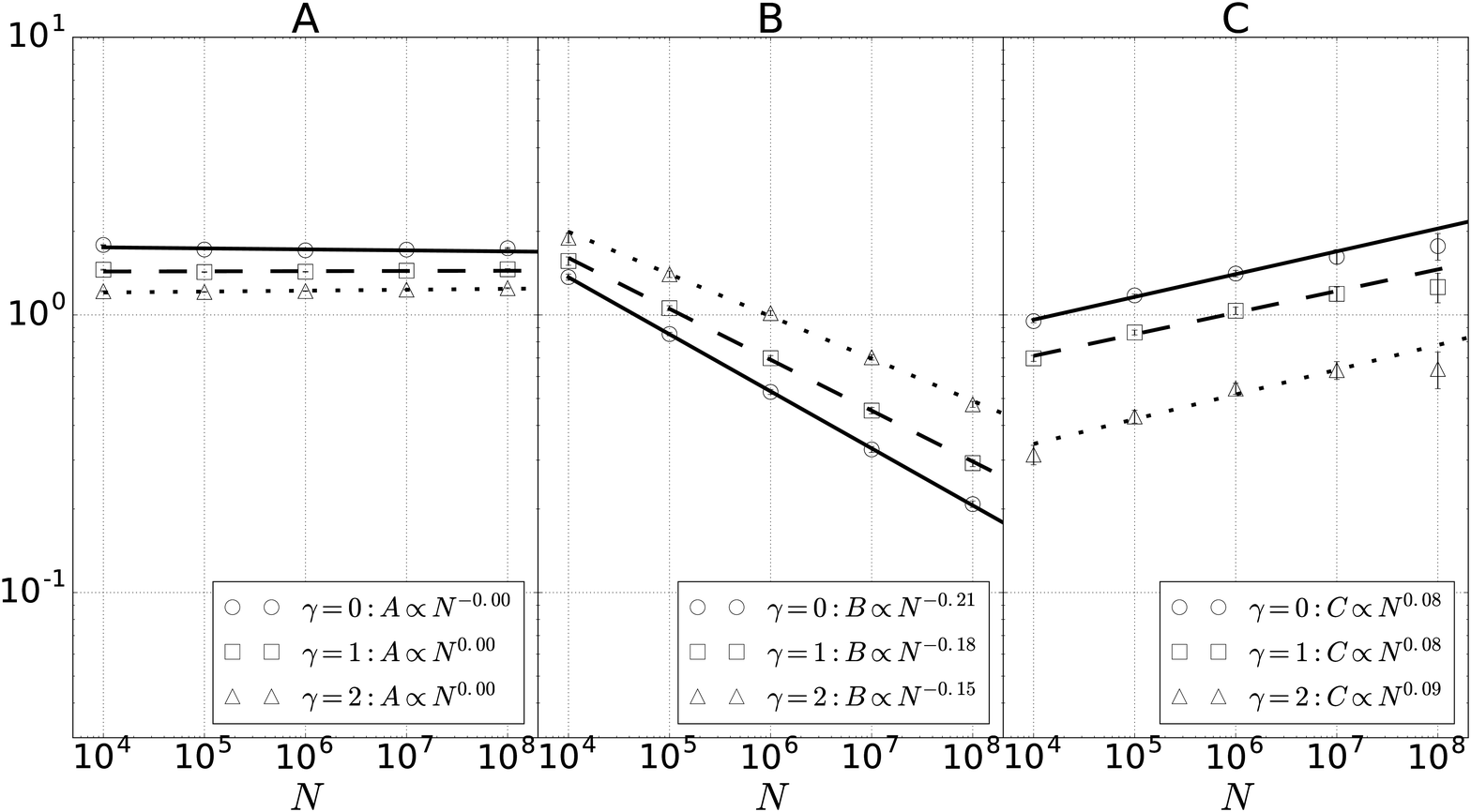}
  \caption{Best fit values of parameters $A$, $B$ and $C$ in
    Eq.~\eqref{eq:fit_delta_S} obtained with data shown in
    Fig.~\ref{fig:S_unif_sph_ellips} (initial uniform
    sphere). Parameters $B$ and $C$ have a power law
    $N$-dependence. For $B$, the $N$-dependence is weaker in models
    with larger $\gamma$, i.e. for potentials hosting larger fractions
    of chaotic orbits.}
  \label{fig:params_unif_sph_ellips}
\end{figure*}

Eq.~\eqref{eq:fit_delta_S} provides a reasonable fit for ensembles of
all sizes $N$ in various models, as shown by solid lines in
Fig.~\ref{fig:S_unif_sph_ellips}. These fits allow us to study the
$N-$dependence of parameters $A$, $B$ and $C$.
Fig. \ref{fig:params_unif_sph_ellips} shows that the parameter $A$ is
nearly constant, whereas $B(N)$ and $C(N)$ can be fitted by power
laws, which are used to predict the behavior for a typical number of
stars in a galaxy, $N=10^{11}$ (dashed lines in
Fig.~\ref{fig:S_unif_sph_ellips}). Since the nature of dark matter,
i.e. its constitution, is still completely unknown, with candidates
ranging from ultralight bosons to massive primordial black holes
\citep[see][]{2018Natur.562...51B}, we do not speculate here about its
number in a typical galaxy.

The parameter $B$ is smaller (and $C$ is larger) for the Perfect
Ellipsoid (${\gamma=0}$, integrable) than for the strong cusp model
(${\gamma=2}$, non-integrable, hosting large fractions of chaotic
orbits, as shown in \S\ref{sec:freq_analysis}). This indicates that
the presence of chaotic orbits anticipates and increases the rate of
entropy production.

We estimate the typical relaxation time as the time $T_{\Delta S/2}$
when the entropy achieves half of its asymptotic value
$A$. Substituting this definition in Eq.\eqref{eq:fit_delta_S}, we
have
\begin{equation}
   \label{eq:T_delta_S_2}
   \frac{T_{\Delta S/2}}{\tau_{cr}} = \sqrt{B^{-2}+C^2}.
\end{equation}

\begin{figure}
  \raggedright
  \includegraphics[width=9.0cm]{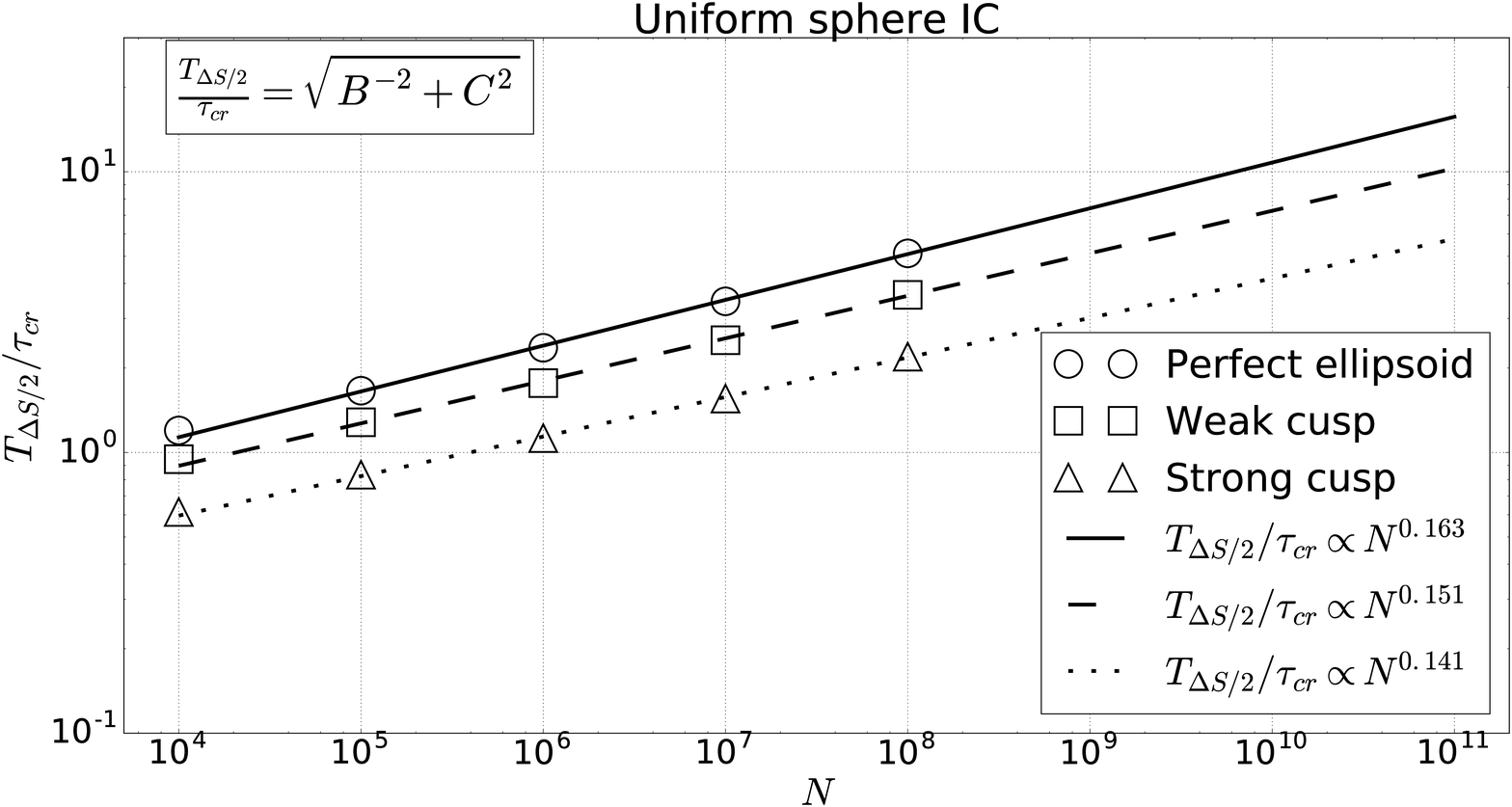}
  \vspace{-0.2cm}
  \caption{Relaxation time, Eq. \eqref{eq:T_delta_S_2}, for the
    uniform sphere initial condition. The data are fitted by power
    laws, $\propto N^{\alpha/d}$, a consequence of the Nyquist-Shannon
    criterion, Eq.~\eqref{eq:nyquist}. For the Perfect Ellipsoid,
    $\alpha\approx 0.98$, in agreement for a bandwidth $K$ growing
    linearly with time for a system in $d=6$ - compare
    Eqs.~\eqref{eq:T_power_law} and \eqref{eq:nyquist}. Chaotic orbits
    produced by the inner cusps accelerate the relaxation and weaken
    its $N$-dependence.}
  \label{fig:T_vs_N_unif_sph_ellips}
\end{figure}

The points in Fig. \ref{fig:T_vs_N_unif_sph_ellips} represent this
quantity, calculated with the best fit values of parameters $B$ and
$C$. These points can be well fitted by power laws
\begin{equation}
  \label{eq:T_power_law}
  T_{\Delta S/2}/\tau_{cr} \propto N^{\alpha/d},
\end{equation}
where $d=6$ is the dimension of the phase-space (the reason for
writing Eq.~\eqref{eq:T_power_law} in this format will be clear
below). These power law fits are shown as black lines in
Fig. \ref{fig:T_vs_N_unif_sph_ellips}. For all models the entropy has
a significant increase after $1-10 \tau_{cr}$, even in the
extrapolation to ${N=10^{11}}$. Note that smaller values of
$T_{\Delta S/2}$ represent an earlier entropy production, and once
more we conclude that the presence of chaotic orbits anticipates it.

In \cite{BeS_2018_2} a simple connection between the entropy evolution
and the Nyquist-Shannon theorem is shown. This theorem states a
one-to-one correspondence between a function in a $d$-dimensional
continuum and a discrete sample of it \emph{if} the number of sampling
points is $N \gtrsim K^d$, where $K$ is the function bandwidth
(i.e. the inverse size of its smallest substructures). Conversely,
given a sample of size $N$ the theorem states that only functions with
\begin{equation}
  \label{eq:nyquist}
  K \lesssim N^{1/d},
\end{equation}
i.e. with structures not too fine, can be uniquely associated to the
sample.

\cite{BeS_2018_2} have shown that the entropy estimated with
Eqs.~\eqref{eq:S_estimate_0}-\eqref{eq:f_NN} is in good agreement with
the Nyquist-Shannon criterion, Eq.~\eqref{eq:nyquist}. As a
consequence, the power law for the relaxation time,
Eq.\eqref{eq:T_power_law}, can be seen as a natural consequence of the
connection with the Nyquist-Shannon theorem. For a system with
phase-space structures evolving in such a way that the bandwidth $K$
of the distribution function grows linearly with time, we would expect
$\alpha/d=1/6 \approx 0.166$. Note the striking value
$\alpha/d \approx 0.163 \rightarrow \alpha\approx 0.98$ obtained for
the Perfect Ellipsoid. For the weak and strong cusp models, the power
laws correspond to $\alpha \approx 0.91$ and $\alpha \approx 0.85$,
respectively (see Fig.~\ref{fig:T_vs_N_unif_sph_ellips}).

For long-range interacting systems in $d=2$,
\cite{2017JSMTE..04.4001P} also found power laws for the
$N$-dependence of the typical time for entropy increase, with
$\alpha = 1$ in an integrable model. This result also fits in the
association with the Nyquist-Shannon criterion, Eq.\eqref{eq:nyquist},
for a bandwidth increasing linearly with time. Interestingly,
\cite{2017JSMTE..04.4001P} also found a weaker $N$-dependence (smaller
$\alpha$) for non-integrable systems, which they interpret as a
consequence of the development of more complex phase-space structures
in the presence of chaotic motion. The weaker $N$-dependence obtained
here for the non-integrable models $\gamma=1,2$
(Fig.\ref{fig:T_vs_N_unif_sph_ellips}) seems in line with this. As
shown by \cite{BeS_2018_2}, this weakening of the $N$-dependence in
non-integrable models is associated to a bandwidth growing (i.e. fine
structures developing) faster than linearly in time for these
non-integrable models.

\subsection{Entropy evolution: Gaussian initial conditions}
\label{sec:entropy_gauss_ic}
\begin{figure*}
  \epsscale{0.85}
  \plotone{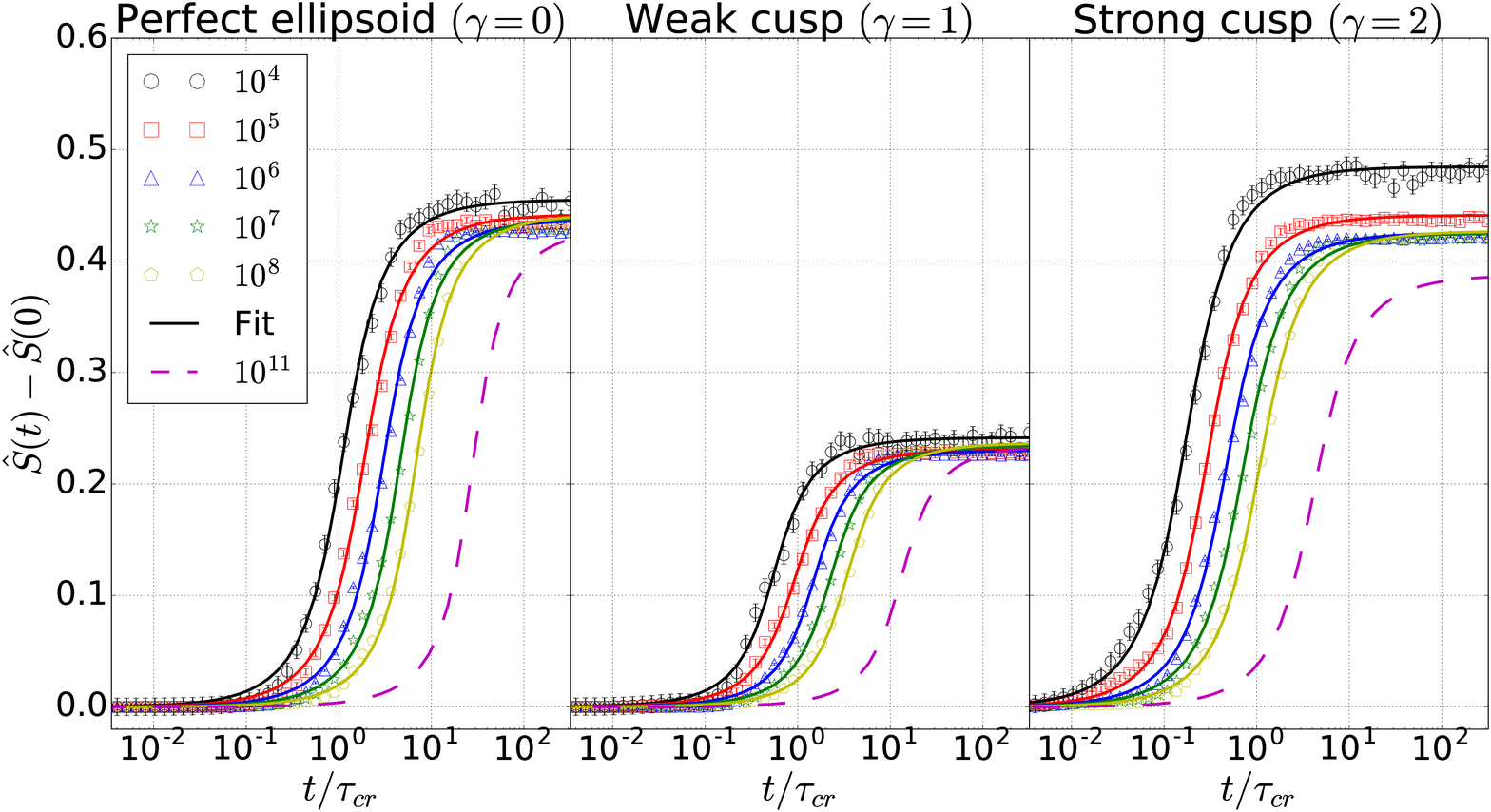}
  \caption{Same as Fig.~\ref{fig:S_unif_sph_ellips}, but using the
    Gaussian initial conditions. The low entropy production for
    $\gamma=1$ can be attributed to the large number of orbits trapped
    by resonances, as shown in Fig.~\ref{fig:freq_map_10k_gauss}.}
  \label{fig:S_gauss_ellips}
\end{figure*}

Fig.~\ref{fig:S_gauss_ellips} shows the entropy evolution for
ensembles of orbits with the Gaussian initial conditions, again in the
different models and for various numbers of particles. The qualitative
behavior is similar to that observed for the uniform sphere initial
condition, with a fast entropy increase, achieving a maximum after
$10-50\tau_{cr}$.

In contrast with the uniform sphere initial condition, the final
entropy amplitude changes non-monotonically from the Perfect Ellipsoid
($\gamma = 0$) to the strong cusp model ($\gamma=2$). While at first
this appears surprising, it is important to remember that the
available phase-space volume and the rate at which orbits fill it
depend on (a) the geometry of the phase-space and (b) the types of
orbits in the ensemble (regular, resonantly trapped or chaotic, and
the degree of chaoticity). In addition, the fraction of resonant
orbits in a system varies with energy
\citep[][]{deibel_etal_11}. Orbits that are trapped around resonances
(either regular resonant or ``sticky'' chaotic orbits) are confined to
a phase space of lower dimensionality than regular or strongly chaotic
orbits.  
As shown in \S~\ref{sec:freq_analysis}, in the weak cusp model
($\gamma =1$) there is a prominent presence of resonantly trapped
orbits. It is interesting to remember that when an orbit is trapped at
or near a resonance it is effectively constrained to a phase-space of
fewer degrees of freedom and therefore represents a restriction in the
phase-space volume explored by the orbit in comparison to the volume
it could have explored given its energy. Thus it is natural that a
large fraction of resonant orbits will produce a smaller amount of
phase mixing and consequently a smaller entropy production, in
comparison to non-resonant orbits\footnote{Interestingly, this
  suggests that, contrary to the common-sense idea that chaos
  introduces disorder, a perturbation in an integrable model producing
  a small amount of chaos can introduce \emph{order} in respect to the
  integrable model.}.

Since the number of resonances increases as one gets deeper in the
potential \citep[see Figs.~12 \& 15 of][]{deibel_etal_11}, and since
Gaussian initial conditions result in more orbits exploring the
phase-space associated with a deeper potential, it is not surprising
that the entropy production is more severely hampered by resonant
trapping. This seems to be the reason for the small entropy production
observed for the Gaussian initial conditions integrated in the weak
cusp model $\gamma=1$, in comparison to the other models -- see
Fig.~\ref{fig:S_gauss_ellips}. We speculate that this non-monotonicity
of the final entropy value as a function of $\gamma$ is not observed
for the initial uniform sphere, in Fig.~\ref{fig:S_unif_sph_ellips},
because a smaller fraction of orbits explore the inner region and
there is a greater entropy production resulting from phase mixing
arising due to this much broader spread of initial conditions (note
the larger values of the final entropy amplitude in this case).

\begin{figure*}
  \epsscale{0.85}
  \plotone{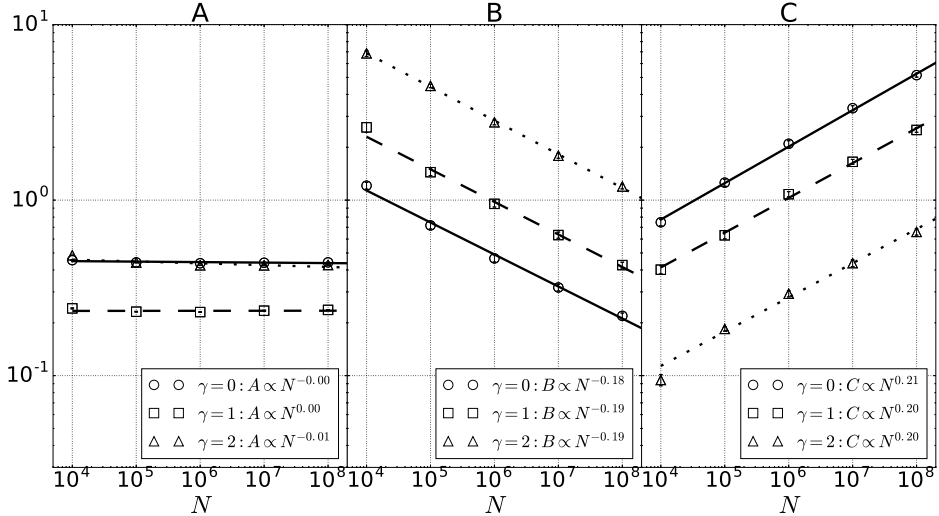}
  \caption{Same as Fig.~\ref{fig:params_unif_sph_ellips}, but using
    the Gaussian initial conditions.}
  \label{fig:params_gauss_ellips}
\end{figure*}

Again, Eq.~\eqref{eq:fit_delta_S} provides a reasonable fit to data --
solid lines in Fig.~\ref{fig:S_gauss_ellips}. The $N$-dependencies of
parameters $A$, $B$ and $C$ are shown in
Fig.~\ref{fig:params_gauss_ellips}. Again parameter $A$ is
approximately constant, while parameters $B(N)$ and $C(N)$ can be
fitted by power laws, similarly to what we obtained for the uniform
sphere initial condition. These power laws are again used to predict
the entropy evolution for $N=10^{11}$ (dashed lines in
Fig.~\ref{fig:S_gauss_ellips}).

\begin{figure}
  \raggedright
  \includegraphics[width=9.0cm]{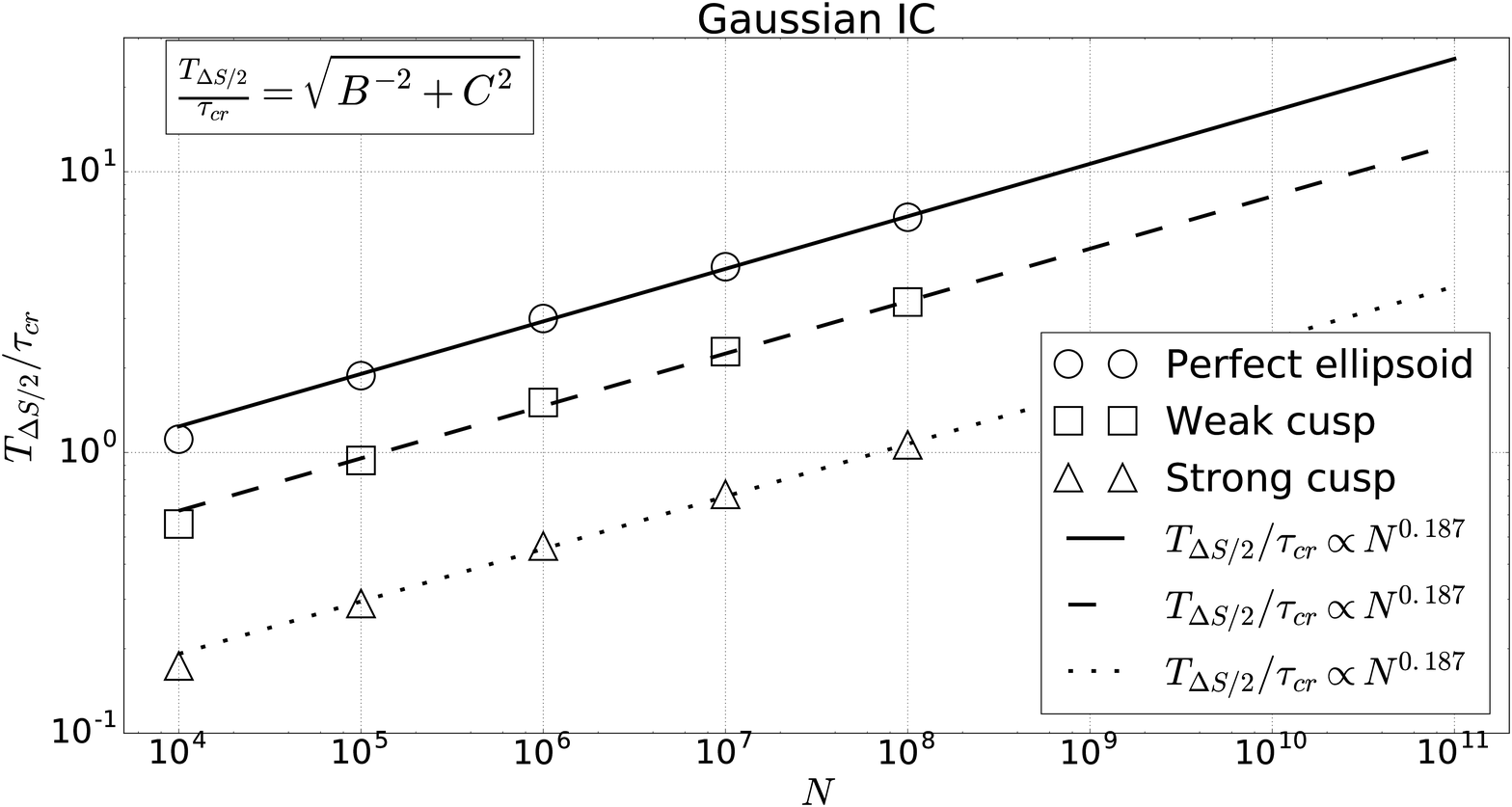}
  \vspace{-0.2cm}
  \caption{Same as Fig.~\ref{fig:T_vs_N_unif_sph_ellips}, but using
    the Gaussian initial conditions. In this case, fitting a power law
    $\propto N^{\alpha/d}$ we obtain $\alpha \approx 1.12$ for all
    three models.}
  \label{fig:T_vs_N_gauss_ellips}
\end{figure}

With the fitted values of parameters $B$ and $C$, we calculate the
relaxation time given by
Eq.~\eqref{eq:T_delta_S_2}. Fig.~\ref{fig:T_vs_N_gauss_ellips} shows
this quantity for the different models ($\gamma=0,1,2$), as well as
power law fits to these data. These power laws,
Eq.~\eqref{eq:T_power_law}, can again be seen as related to the
Nyquist-Shannon criterion, Eq. \eqref{eq:nyquist}. In this case the
slope implies $\alpha\approx 1.12$, which is reasonably different from
the value obtained for the uniform sphere initial condition, expected
for a linear time growth of the bandwidth ($\alpha \approx 1$).
Besides that, in this case the slope $\alpha$ is the same for the
different models ($\gamma=0,1,2$). The exact reason for this different
behavior is not clear, but it seems to be related to the narrowness of
the Gaussian initial conditions, in comparison to the much broader
uniform sphere. On the other hand, similarly to the uniform sphere,
the presence of chaotic orbits seems to accelerate the entropy
production (smaller values of $T_{\Delta S/2}/\tau_{cr}$).

Even though there seems to be some differences, the entropy evolution
is still similar in the two sets of initial conditions. Specifically,
in both initial conditions we conclude that the entropy has a
significant entropy increase after $1-10\tau_{cr}$ for a broad range
in $N$ and that this relaxation time scales as $T\propto N^{\alpha/d}$
with $\alpha$ not too different from $1$, even in non-integrable
models. It is interesting to contrast this conclusion with the results
obtained by \cite{1998MNRAS.301..960K}, where the orbit integration of
initially very localized ensembles gives rise to macroscopic evolution
occurring in very different rates for phase mixing in comparison to
chaotic mixing: the former was observed to evolve with a linear rate
and the latter with an exponential rate. However, as pointed out by
\cite{1999PASP..111..129M}, phase mixing of non-localized initial
conditions can be much faster, and the similarity of entropy evolution
observed in Figs. \ref{fig:S_unif_sph_ellips} and
\ref{fig:S_gauss_ellips} seems to be in accordance with this
observation.

\section{Meaning of the entropy estimator}
\label{sec:meaning_estimator}

Since they are applied to finite, discrete samples, the estimators
used in this work do not (directly) depend on the distribution
function at interparticle phase-space positions, but only on
(estimates of) $f$ at the position of each particle, using the
information available in its neighborhood. One could argue then that
these estimators involve some kind of coarse-graining procedure,
meaning that we are averaging the true underlying distribution
function in a finite region and thus loosing information that could be
completely recovered only in the continuous limit. Note that the same
argument is normally used to deny the phenomenon of macroscopic time
irreversibility in general. According to this argument, the entropy
increase observed here would not represent a real physical effect, but
it would be the mere consequence of information loss in the
measurement (estimation) process. The time irreversibility is thus
relegated to the status of a subjective, non-physical effect.

This solution, however, seems deeply unsatisfactory, since macroscopic
irreversibility is an objective fact and cannot depend on our
measurement precision \citep[see][]{1965AmJPh..33..391J}. In
particular, a galaxy collapses and forms in one time direction (and
not the other) independent of any subjective observational
coarse-graining.

In \cite{BeS_2018_2} it is shown that the entropy estimator used here
is in agreement with the Nyquist-Shannon criterion, i.e. that it
coincides with the entropy of the \emph{assumed} distribution function
in the continuum, as far as its bandwidth (in frequency/Fourier space)
is not too large -- see Eq. \eqref{eq:nyquist}. On the other hand,
given a discrete sample, any assumed distribution function with
bandwidth larger than this limit value has structures too fine to be
realized by the sample and can be seen as an \textit{information
  input}. Let us remember that what is given a priori in real systems
is a discrete sample, the continuum limit being a mathematical
abstraction which can or cannot represent a good approximation in the
description of the phenomenon. Once this extrapolation to the
continuum is done, at a first sight it seems that the entropy
estimators produce an information loss. However, these estimators only
recover the information we actually have, which is contained in the
sample and not in any extrapolation to the continuum.

As noted by \cite{NYAS:NYAS28}, gravitational systems are
intrinsically inhomogeneous and one important question is whether the
number of constituent particles (or stars) suffices for the continuum
limit $(N\rightarrow\infty)$ to accurately describe the evolution of
the system over the timescales of interest -- see also
\cite{doi:10.1093/mnras/sty1728}. The results of
\S~\ref{sec:ellipsoid} show that for a finite-$N$ gravitational system
in an external potential the entropy increases significantly after a
typical time
\begin{equation}
  \label{eq:t_relax}
  T_{\Delta/2}/\tau_{cr} \approx 0.1 N^{1/6}.
\end{equation}
Thus, even values as large as $N=10^{11}$, representative of the
number of stars in real galaxies, are not enough for the continuum
limit to be a valid approximation after a few crossing times -- see
Figs.~\ref{fig:S_unif_sph_ellips} and \ref{fig:S_gauss_ellips}. This
constitutes a real (and fast) relaxation phenomenon and the entropy
estimator captures the time irreversibility associated to it. We call
it \emph{(collisionless) discreteness-driven relaxation}, in contrast
to the \emph{violent relaxation} proposed by \cite{LyndenBell_1967}.

\section{Collisionless relaxation and Vlasov equation}
\label{sec:meaning_vlasov}

Having shown that the observed entropy increase is a real effect
associated to the intrinsic discreteness of gravitating systems, we
move on to the discussion of its kinetic description. This early
entropy evolution is similar to what was observed for $N$-body
simulations in \citetalias{2017ApJ...846..125B}, in a regime
associated to violent relaxation. This phenomenon is traditionally
assumed to be described by the Vlasov equation, Eq.~\eqref{eq:vlasov}.

The fundamental problem with this traditional view is that the Vlasov
equation is time-reversible and implies entropy conservation
\citep[see][]{Tremaine_Henon_Lynden_Bell_1986}. This already suggests
that this equation does not contain the physical ingredients necessary
to provide a kinetic description of the fast relaxation of
collisionless systems. In fact, \cite{1990ApJ...351..104K} has shown
that the Vlasov-Poisson equation constitutes a Hamiltonian system with
the distribution function being a dynamical variable defined in an
infinite-dimensional phase-space. In other words, this suggests that
the description provided by the Vlasov equation, at least as it is
used in the context of self-gravitating systems, is essentially
identical to the microscopic description, i.e. at the level of single
trajectories, provided by Classical Mechanics, with possible technical
advantages with respect to $N$-body simulations
\citep[see][]{2013ApJ...762..116Y, 2015MNRAS.450.3724C,
  2016MNRAS.455.1115H}. However, as a fundamental disadvantage, the
Vlasov equation, as any purely mechanical approach, is not able to
describe phenomena that only emerge at a macroscopic level, mainly the
time irreversibility expressed by the 2nd law of Thermodynamics, which
can be considered ``\textit{one of the most perfect laws in physics}''
\citep[][]{LIEB19991} and has never been faulted by reproducible
experiments.

Additionally, all the rigorous mathematical results on the Vlasov
equation for a given finite-$N$ system are only able to prove its
validity for a finite time -- see \cite{BoePick, LaPick}. It is thus
not surprising that this property is numerically verified, as also
reported by \cite{10.1088/1751-8121/aaea0c} (appeared during the
revision process of the current paper), where the polynomial behavior
in $N$ for the validity time of the Vlasov equation is also
independently emphasized, albeit without any conceptual explanation
nor a precise study on the polynomial law \citep[as done
by][]{BeS_2018_2}. At this point, recall that the convergence of the
effective distribution function to its thermodynamic limit
$N\rightarrow \infty $ is \emph{not uniform} with respect to time, in
all mathematically rigorous results on the validity of the Vlasov
equation. That is, even if one can approximate arbitrarily well the
effective distribution for a $N$-body system via a solution of the
Vlasov equation, for some fixed sufficiently large $N_{0}$ and on a
\emph{fixed} time interval, say $[0,T]$, this function would not
anymore, in general, well approximate the effective distribution of
the finite system for times $t>2T$ (the new $N$ necessary to provide a
good approximation on $[0,2T]$ is usually $\gg N_{0}$). In other
words, everything depends on the involved time- and $N$-scales of the
system under consideration. If one studies $1~\mathrm{l}$ of liquid,
corresponding to $N=10^{24}$ (cf. the Avogadro constant), for times of
order of years, the thermodynamic limit $N\rightarrow \infty $ seems
to be perfectly justified, at least empirically. This is much more
questionable for gravitating systems such as galaxies, with typically
$N\lesssim 10^{11}$ and ages of $\approx 10^{10}$ years, unless one
dogmatically postulates that such objects are well-described by the
limit $N\rightarrow\infty$.  In fact, criticizing such a postulate is
a central point here, as well as in \citetalias{2017ApJ...846..125B}
and \cite{BeS_2018_2}. What is more, even in the continuous limit
$N\rightarrow\infty $, the very notion of large-time convergence of a
sequence of distribution functions developing rapidly varying
structures (filaments) with the time evolution is a non-trivial
conceptual point: such sequences cannot converge in the point-wise
sense and the so-called weak convergence is a more natural notion in
this situation, as pointed out by \cite{mouhot2011}. This type of
convergence means, roughly, that structures that get arbitrarily fine
in the limit must be ``averaged out'' in order to obtain a
well-defined limiting distribution. It is well-known that the entropy
is \emph{not} continuous with respect to such a \emph{weak}
convergence and therefore, an entropy production for the finite system
is clearly not in contradiction with the constant-entropy of the
Vlasov equation at any fixed, finite time. Our approach \citep[see in
particular][]{BeS_2018_2} sheds some light on this question, by
providing a quantitative criterion to objectively evaluate the
``collapse'' of fine structures of distributions of particles of
macroscopic systems, at fixed (finite) $N$.


Let us stress that although we evolve the orbits in a fixed potential,
we do so to eliminate the possibility that a time-dependence
\citep[which is invoked in violent relaxation scenario proposed
by][]{LyndenBell_1967} could be affecting the relaxation of the
ensemble. This does not mean that in practice the potential
fluctuations do not occur or that they do not have any effect on the
evolution of the ensemble. Our goal in using a fixed potential is to
demonstrate that this discreteness-driven relaxation occurs whenever
the phase-space distribution of an ensemble is not in dynamical
equilibrium with the potential. By using a fixed potential and an
ensemble whose phase-space distribution is not self-consistent with
the potential, we show that this alone will cause this relaxation to
occur, on short timescales. The potential fluctuations that accompany
the formation of real systems will likely accelerate this process.

Thus, even though the early collisionless relaxation of
self-gravitating $N$-body systems must be much more complex than the
orbit integration explored in this work, our results shed light on the
role of its main ingredients: a finite number of particles and the
anticipation/acceleration of entropy production in the presence of
chaotic motion. Since this early relaxation (together with a
time-varying potential) is expected to host a large amount of chaos
\citep[see][]{2003MNRAS.341..927K}, the entropy increase in real
$N$-body systems can be even faster, invalidating even sooner the
applicability of the Vlasov equation as a description of the
macroscopic evolution. Note that this occurs even in absence of any
collisional relaxation, and this is why, as already mentioned in
\citetalias{2017ApJ...846..125B}, we prefer not to call
Eq.~\eqref{eq:vlasov} the ``Collisionless Boltzmann equation'', as
suggested by \cite{1982A&A...114..211H}.

The timescale in Eq.~\eqref{eq:t_relax}, derived with reference to the
Nyquist-Shannon theorem by \cite{BeS_2018_2}, can be seen as a
theoretical explanation for the fast character of the collisionless
relaxation of $N$-body gravitating systems.

\section{Conclusions}
\label{sec:conclusions}
In this work, we integrate ensembles of orbits in fixed external
gravitational potentials, studying the entropy evolution of the
ensemble. While this is a much simpler problem than the $N$-body
simulations investigated in \citetalias{2017ApJ...846..125B}, the
current analysis capture the essential ingredients for the time
irreversibility of its early collisionless relaxation. The conclusions
are summarized below:

\begin{itemize}
\item The orbit integration in the Harmonic Potential shows that the
  entropy estimator is perfectly able to recover the macroscopic time
  reversibility in exceptional cases where it is present. This
  indicates that this estimator does not introduce any artificial
  entropy increase, while making clear the difference between
  microscopic time reversibility (always present in the equations of
  motion for each particle) and macroscopic time reversibility,
  present in this potential but not in general models.

\item Integration in the Plummer potential shows that, for a
  non-self-consistent initial condition, the entropy has a fast
  increase (due to phase mixing), achieving a maximum after
  $\sim 10\tau_{cr}$ for ${N=10^6}$. This macroscopic irreversibility
  occurs despite the potential being static and spherical
  (integrable). On the other hand, the estimator correctly captures
  the entropy conservation associated to a self-consistent
  (i.e. stationary) sample of the model. This shows again that this
  estimator does not introduce any artificial entropy increase,
  behaving in accordance with the 2nd law of Thermodynamics.

\item We also investigate a triaxial model whose density profile is
  $\rho(r) \propto r^{-4}$ in the external regions and has a free
  parameter $\gamma$ for the inner slope. For $\gamma=0$, this model
  reduces to the Perfect Ellipsoid, which is integrable. Larger values
  of $\gamma$ (inner cusps) generate increasing fractions of chaotic
  orbits, as shown by means of a frequency analysis. This analysis
  also shows that a large fraction of orbits are resonantly trapped in
  the weak cusp model ($\gamma=1$), producing heavily populated
  resonance lines, which are destroyed in the strong cusp model
  ($\gamma=2$).

\item We derive a typical relaxation time that scales as
  $T/\tau_{cr} \propto N^{1/6}$ for an initial ensemble sampling a
  uniform sphere (in positions and velocities) evolved in an
  integrable potential. Similar $N$-dependencies are found for a
  different initial condition and for non-integrable models.

\item The presence of chaotic orbits seems to accelerate the entropy
  production (see Figs.~\ref{fig:T_vs_N_unif_sph_ellips} and
  \ref{fig:T_vs_N_gauss_ellips}), as found by
  \cite{2017JSMTE..04.4001P} for $d=2$.

\item This power law $N$-dependence of the typical relaxation
  timescale can be seen as a natural consequence of the
  Nyquist-Shannon criterion, as pointed out by \cite{BeS_2018_2}. In
  this way, the key point for macroscopic time irreversibility is the
  fact that the system is discrete, i.e. composed of a finite number
  $N$ of elements, regardless of the presence of chaotic motion or a
  time-dependent collective potential, in line with
  \cite{1993PhyA..194....1L}'s ideas.

\item This connection with the Nyquist-Shannon criterion makes clear
  the objectivity of the relaxation and entropy increase, without need
  of the subjective idea of information loss due to coarse-graining.

\item The derived timescale, Eq.~\eqref{eq:t_relax}, can be seen as an
  upper limit for the timescale of the collisionless relaxation of
  real collapsing $N$-body systems, since the collapse (together with
  a time-varying potential) is expected to produce a large amount of
  chaotic orbits \citep[see][]{2003MNRAS.341..927K}, which would tend
  to accelerate the entropy production.

\item Reinforcing the conclusion drawn in
  \citetalias{2017ApJ...846..125B}, our results indicate that the
  Vlasov equation is not able to provide a kinetic description
  (i.e. in a macroscopic level) of the early collisionless relaxation
  of gravitating systems.

\end{itemize}

Some improvements for future work would be a study of the
$N$-dependence of the entropy evolution in self-gravitating $N$-body
simulations and the prediction of the final entropy value, which could
involve identifying the correct constraints in a maximization
procedure \citep[see][for recent attempts]{Hjorth_Williams_2010_I,
  Pontzen_Governato_2013, Beraldo_Lima_Sodre_Perez_2014}.

\section*{Acknowledgements}
We thank E. Vasiliev for a careful reading and comments. This work has
made use of the computing facilities of the Laboratory of
Astroinformatics (IAG/USP, NAT/Unicsul), whose purchase was made
possible by the Brazilian agency FAPESP (2009/54006-4) and the
INCT-A. LBeS is supported by FAPESP (2014/23751-4 and
2017-01421-0). WdSP is supported by CNPq (308337/2017-4). MV
acknowledges support from HST-AR-13890.001, NSF award AST-1515001,
NASA-ATP award NNX15AK79G. LSJ is supported by FAPESP (2017/25620-2)
and CNPq. JBB is supported by FAPESP (2017/22340-9), by the Basque
Government (IT641-13 and BERC 2018-2022 program), and by the Spanish
Ministry of Science, Innovation and Universities: BCAM Severo Ochoa
accreditation SEV-2017-0718, MTM2017-82160-C2-2-P. This paper made use
of Agama \citep{2018MNRAS.tmp.2556V}, ANN
\citep{Arya:1998:OAA:293347.293348}, GSL, matplotlib
\citep{Hunter:2007}, numpy \citep{Walt:2011:NAS:1957373.1957466} and
scipy \citep{scipy}.

\bibliography{/Users/lbs/refs_lbs}

\newcommand{\noop}[1]{}
\begin{thebibliography}{}
\expandafter\ifx\csname natexlab\endcsname\relax\def\natexlab#1{#1}\fi

\bibitem[{{Aarseth} {et~al.}(1974){Aarseth}, {Henon}, \&
  {Wielen}}]{1974A&A....37..183A}
{Aarseth}, S.~J., {Henon}, M., \& {Wielen}, R. 1974, AAp, 37, 183

\bibitem[{Anderson(1972)}]{Anderson393}
Anderson, P.~W. 1972, Science, 177, 393

\bibitem[{Arya {et~al.}(1998)Arya, Mount, Netanyahu, Silverman, \&
  Wu}]{Arya:1998:OAA:293347.293348}
Arya, S., Mount, D.~M., Netanyahu, N.~S., Silverman, R., \& Wu, A.~Y. 1998, J.
  ACM, 45, 891

\bibitem[{Beirlant {et~al.}(1997)Beirlant, Dudewicz, Gy\"{o}rfi, \& {Van Der
  Meulen}}]{Beirlant1997a}
Beirlant, J., Dudewicz, E.~J., Gy\"{o}rfi, L., \& {Van Der Meulen}, E.~C. 1997,
  International Journal of Mathematical and Statistical Sciences, 6, 17

\bibitem[{{Beraldo e Silva} {et~al.}(2017){Beraldo e Silva}, {de Siqueira
  Pedra}, {Sodr{\'e}}, {Perico}, \& {Lima}}]{2017ApJ...846..125B}
{Beraldo e Silva}, L., {de Siqueira Pedra}, W., {Sodr{\'e}}, L., {Perico},
  E.~L.~D., \& {Lima}, M. 2017, ApJ, 846, 125

\bibitem[{{Beraldo e Silva} {et~al.}({2018} submitted){Beraldo e Silva}, {de
  Siqueira Pedra}, \& Valluri}]{BeS_2018_2}
{Beraldo e Silva}, L., {de Siqueira Pedra}, W., \& Valluri, M. {2018}
  submitted, ApJ

\bibitem[{Beraldo~e Silva {et~al.}(2014)Beraldo~e Silva, Lima, Sodr\'e, \&
  Perez}]{Beraldo_Lima_Sodre_Perez_2014}
Beraldo~e Silva, L., Lima, M., Sodr\'e, L., \& Perez, J. 2014, Phys. Rev. D,
  90, 123004

\bibitem[{{Bertone} \& {Tait}(2018)}]{2018Natur.562...51B}
{Bertone}, G., \& {Tait}, T.~M.~P. 2018, Nature, 562, 51

\bibitem[{Biau \& Devroye(2015)}]{biau2015lectures}
Biau, G., \& Devroye, L. 2015, Lectures on the Nearest Neighbor Method,
  Springer Series in the Data Sciences (Springer International Publishing)

\bibitem[{{Binney} \& {Spergel}(1982)}]{1982ApJ...252..308B}
{Binney}, J., \& {Spergel}, D. 1982, ApJ, 252, 308

\bibitem[{Binney \& Tremaine(2008)}]{Binney_2008}
Binney, J., \& Tremaine, S. 2008, Galactic Dynamics - Second Edition (Princeton
  University Press)

\bibitem[{Boers \& Pickl(2016)}]{BoePick}
Boers, N., \& Pickl, P. 2016, JSP, 164, 1

\bibitem[{Boltzmann(1974)}]{book:969643}
Boltzmann, L. 1974, Theoretical Physics and Philosophical Problems: Selected
  Writings, 1st edn., ed. B.~McGuinness, Vienna Circle Collection 5 (Springer
  Netherlands)

\bibitem[{Bru \& {de Siqueira Pedra}(2015)}]{doi:10.1142/S0218202515500566}
Bru, J.-B., \& {de Siqueira Pedra}, W. 2015, Mathematical Models and Methods in
  Applied Sciences, 25, 2587

\bibitem[{Cercignani(1988)}]{cercignani1988boltzmann}
Cercignani, C. 1988, The Boltzmann Equation and Its Applications, Applied
  mathematical sciences No. v. 67 (Springer-Verlag)

\bibitem[{Chirikov(1979)}]{CHIRIKOV1979263}
Chirikov, B.~V. 1979, Physics Reports, 52, 263

\bibitem[{{Colombi} {et~al.}(2015){Colombi}, {Sousbie}, {Peirani}, {Plum}, \&
  {Suto}}]{2015MNRAS.450.3724C}
{Colombi}, S., {Sousbie}, T., {Peirani}, S., {Plum}, G., \& {Suto}, Y. 2015,
  MNRAS, 450, 3724

\bibitem[{{de Zeeuw}(1985)}]{1985MNRAS.216..273D}
{de Zeeuw}, T. 1985, MNRAS, 216, 273

\bibitem[{{Dehnen}(1993)}]{1993MNRAS.265..250D}
{Dehnen}, W. 1993, MNRAS, 265, 250

\bibitem[{{Deibel} {et~al.}(2011){Deibel}, {Valluri}, \&
  {Merritt}}]{deibel_etal_11}
{Deibel}, A.~T., {Valluri}, M., \& {Merritt}, D. 2011, ApJ, 728, 128

\bibitem[{Dobrushin(1979)}]{Dobr}
Dobrushin, R.~L. 1979, FAIA, 13, 115

\bibitem[{Farias {et~al.}(2018)Farias, Pakter, \&
  Levin}]{10.1088/1751-8121/aaea0c}
Farias, C. A.~F., Pakter, R., \& Levin, Y. 2018, JPA: Math. and Theor.

\bibitem[{Ford(1975)}]{cohen1975fundamental}
Ford, J. 1975, Fundamental problems in statistical mechanics III (North-Holland
  Pub. Co.)

\bibitem[{Friedman {et~al.}(1977)Friedman, Bentley, \&
  Finkel}]{Friedman:1977:AFB:355744.355745}
Friedman, J.~H., Bentley, J.~L., \& Finkel, R.~A. 1977, ACM Trans. Math.
  Softw., 3, 209

\bibitem[{{Hahn} \& {Angulo}(2016)}]{2016MNRAS.455.1115H}
{Hahn}, O., \& {Angulo}, R.~E. 2016, MNRAS, 455, 1115

\bibitem[{{Hemsendorf} \& {Merritt}(2002)}]{2002ApJ...580..606H}
{Hemsendorf}, M., \& {Merritt}, D. 2002, ApJ, 580, 606

\bibitem[{{H\'enon}(1982)}]{1982A&A...114..211H}
{H\'enon}, M. 1982, AAP, 114, 211

\bibitem[{Hjorth \& Williams(2010)}]{Hjorth_Williams_2010_I}
Hjorth, J., \& Williams, L.~R. 2010, ApJ, 722, 851

\bibitem[{Hunter(2007)}]{Hunter:2007}
Hunter, J.~D. 2007, Computing In Science \& Engineering, 9, 90

\bibitem[{{Jaynes}(1965)}]{1965AmJPh..33..391J}
{Jaynes}, E.~T. 1965, Am. J. Phys., 33, 391

\bibitem[{Joe(1989)}]{Joe1989}
Joe, H. 1989, AISM, 41, 683

\bibitem[{Jones {et~al.}(2001--)Jones, Oliphant, Peterson, {et~al.}}]{scipy}
Jones, E., Oliphant, T., Peterson, P., {et~al.} 2001--, {SciPy}: Open source
  scientific tools for {Python}, ,

\bibitem[{{Kandrup}(1990)}]{1990ApJ...351..104K}
{Kandrup}, H.~E. 1990, ApJ, 351, 104

\bibitem[{Kandrup(1998)}]{NYAS:NYAS28}
Kandrup, H.~E. 1998, Annals NY Academy of Sciences, 848, 28

\bibitem[{{Kandrup}(1998)}]{1998MNRAS.301..960K}
{Kandrup}, H.~E. 1998, MNRAS, 301, 960

\bibitem[{{Kandrup} {et~al.}(1993){Kandrup}, {Mahon}, \&
  {Smith}}]{Kandrup_1993}
{Kandrup}, H.~E., {Mahon}, M.~E., \& {Smith}, Jr., H. 1993, Astronomy and
  Astrophysics, 271, 440

\bibitem[{{Kandrup} {et~al.}(2003){Kandrup}, {Vass}, \&
  {Sideris}}]{2003MNRAS.341..927K}
{Kandrup}, H.~E., {Vass}, I.~M., \& {Sideris}, I.~V. 2003, MNRAS, 341, 927

\bibitem[{{Krylov} {et~al.}(1979){Krylov}, {Migdal}, {Sinai}, \&
  {Zeeman}}]{1979wfst.book.....K}
{Krylov}, N.~S., {Migdal}, A.~B., {Sinai}, Y.~G., \& {Zeeman}, Y.~L. 1979,
  {Works on the Foundations of Statistical Physics by Nikolai Sergeevich
  Krylov} (Princeton University Press)

\bibitem[{{Laskar}(1990)}]{1990Icar...88..266L}
{Laskar}, J. 1990, Icarus, 88, 266

\bibitem[{{Lazarovici} \& {Pickl}(2017)}]{LaPick}
{Lazarovici}, D., \& {Pickl}, P. 2017, Arch. Rat. Mech. Anal.,
  doi:10.1007/s00205-017-1125-0

\bibitem[{{Lebowitz}(1993)}]{1993PhyA..194....1L}
{Lebowitz}, J.~L. 1993, Physica A, 194, 1

\bibitem[{{Lebowitz}(1999)}]{1999PhyA..263..516L}
---. 1999, Physica A Statistical Mechanics and its Applications, 263, 516

\bibitem[{Lebowitz(2007)}]{lebowitz2007time}
Lebowitz, J.~L. 2007, Boltzmann’s Legacy, 63

\bibitem[{Leonenko {et~al.}(2008)Leonenko, Pronzato, \& Savani}]{Leonenko_2008}
Leonenko, N., Pronzato, L., \& Savani, V. 2008, TATRA MT. MATH. PUBL., 39, 265

\bibitem[{Lichtenberg \& Lieberman(1992)}]{Lichtenberg}
Lichtenberg, A.~J., \& Lieberman, M.~A. 1992, Regular and Chaotic Dynamics, 2nd
  edn., Applied Mathematical Sciences No.~38 (New York, NY: Springer-Verlag)

\bibitem[{Lieb \& Yngvason(1999)}]{LIEB19991}
Lieb, E.~H., \& Yngvason, J. 1999, Physics Reports, 310, 1

\bibitem[{Ludlow {et~al.}(2011)Ludlow, Navarro, White, Boylan-Kolchin,
  Springel, Jenkins, \& Frenk}]{doi:10.1111/j.1365-2966.2011.19008.x}
Ludlow, A.~D., Navarro, J.~F., White, S. D.~M., {et~al.} 2011, MNRAS, 415, 3895

\bibitem[{Lynden-Bell(1967)}]{LyndenBell_1967}
Lynden-Bell, D. 1967, MNRAS, 136, 101

\bibitem[{{May} \& {van Albada}(1984)}]{1984MNRAS.209...15M}
{May}, A., \& {van Albada}, T.~S. 1984, MNRAS, 209, 15

\bibitem[{{McGlynn}(1984)}]{1984ApJ...281...13M}
{McGlynn}, T.~A. 1984, ApJ, 281, 13

\bibitem[{{Merritt}(1999)}]{1999PASP..111..129M}
{Merritt}, D. 1999, PASP, 111, 129

\bibitem[{{Merritt}(2005)}]{2005NYASA1045....3M}
---. 2005, Annals NY Acad. of Sciences, 1045, 3

\bibitem[{{Merritt} \& {Valluri}(1996)}]{Merritt_1996_2}
{Merritt}, D., \& {Valluri}, M. 1996, ApJ, 471, 82

\bibitem[{{Miller}(1964)}]{1964ApJ...140..250M}
{Miller}, R.~H. 1964, ApJ, 140, 250

\bibitem[{Mouhot \& Villani(2011)}]{mouhot2011}
Mouhot, C., \& Villani, C. 2011, Acta Math., 207, 29

\bibitem[{{Navarro} {et~al.}(1997){Navarro}, {Frenk}, \& {White}}]{NFW_1997}
{Navarro}, J.~F., {Frenk}, C.~S., \& {White}, S.~D.~M. 1997, ApJ, 490, 493

\bibitem[{Navarro {et~al.}(2004)Navarro, Hayashi, Power,
  {et~al.}}]{Navarro_2004}
Navarro, J.~F., Hayashi, E., Power, C., {et~al.} 2004, MNRAS, 349, 1039

\bibitem[{{Pakter} \& {Levin}(2017)}]{2017JSMTE..04.4001P}
{Pakter}, R., \& {Levin}, Y. 2017, Journal of Statistical Mechanics: Theory and
  Experiment, 4, 044001

\bibitem[{Pe\~narrubia(2013)}]{doi:10.1093/mnras/stt935}
Pe\~narrubia, J. 2013, MNRAS, 433, 2576

\bibitem[{{Pontzen} \& {Governato}(2013)}]{Pontzen_Governato_2013}
{Pontzen}, A., \& {Governato}, F. 2013, MNRAS, 430, 121

\bibitem[{{Price-Whelan} {et~al.}(2016){Price-Whelan}, {Johnston}, {Valluri},
  {Pearson}, {K{\"u}pper}, \& {Hogg}}]{2016MNRAS.455.1079P}
{Price-Whelan}, A.~M., {Johnston}, K.~V., {Valluri}, M., {et~al.} 2016, MNRAS,
  455, 1079

\bibitem[{{Prigogine}(1999)}]{1999PhyA..263..528P}
{Prigogine}, I. 1999, Physica A Statistical Mechanics and its Applications,
  263, 528

\bibitem[{Romero \& Ascasibar(2018)}]{doi:10.1093/mnras/sty1728}
Romero, M., \& Ascasibar, Y. 2018, MNRAS, sty1728

\bibitem[{{Sharma} \& {Steinmetz}(2006)}]{2006MNRAS.373.1293S}
{Sharma}, S., \& {Steinmetz}, M. 2006, MNRAS, 373, 1293

\bibitem[{{Tremaine} {et~al.}(1986){Tremaine}, {H\'enon}, \&
  {Lynden-Bell}}]{Tremaine_Henon_Lynden_Bell_1986}
{Tremaine}, S., {H\'enon}, M., \& {Lynden-Bell}, D. 1986, MNRAS, 219, 285

\bibitem[{Uhlenbeck(1973)}]{Uhlenbeck1973}
Uhlenbeck, G.~E. 1973, Problems of Statistical Physics, ed. J.~Mehra
  (Dordrecht: Springer Netherlands), 501--513

\bibitem[{Umetsu {et~al.}(2011)Umetsu, Broadhurst, Zitrin, Medezinski, Coe, \&
  Postman}]{Umetsu_2011_b}
Umetsu, K., Broadhurst, T., Zitrin, A., {et~al.} 2011, ApJ, 738, 41

\bibitem[{{Valluri} \& {Merritt}(1998)}]{1998ApJ...506..686V}
{Valluri}, M., \& {Merritt}, D. 1998, ApJ, 506, 686

\bibitem[{{Valluri} {et~al.}(2007){Valluri}, {Vass}, {Kazantzidis}, {Kravtsov},
  \& {Bohn}}]{2007ApJ...658..731V}
{Valluri}, M., {Vass}, I.~M., {Kazantzidis}, S., {Kravtsov}, A.~V., \& {Bohn},
  C.~L. 2007, ApJ, 658, 731

\bibitem[{{van Albada}(1982)}]{1982MNRAS.201..939V}
{van Albada}, T.~S. 1982, MNRAS, 201, 939

\bibitem[{{Vasiliev}(2019)}]{2018MNRAS.tmp.2556V}
{Vasiliev}, E. 2019, MNRAS, 482, 1525

\bibitem[{Walt {et~al.}(2011)Walt, Colbert, \&
  Varoquaux}]{Walt:2011:NAS:1957373.1957466}
Walt, S. v.~d., Colbert, S.~C., \& Varoquaux, G. 2011, Computing in Science and
  Engg., 13, 22

\bibitem[{{Yoshikawa} {et~al.}(2013){Yoshikawa}, {Yoshida}, \&
  {Umemura}}]{2013ApJ...762..116Y}
{Yoshikawa}, K., {Yoshida}, N., \& {Umemura}, M. 2013, APJ, 762, 116

\end{thebibliography}

\appendix
\section{$N$-dependence of entropy estimator errors}
\label{sec:errors}
\begin{figure}
  \raggedright
  \includegraphics[width=9.5cm]{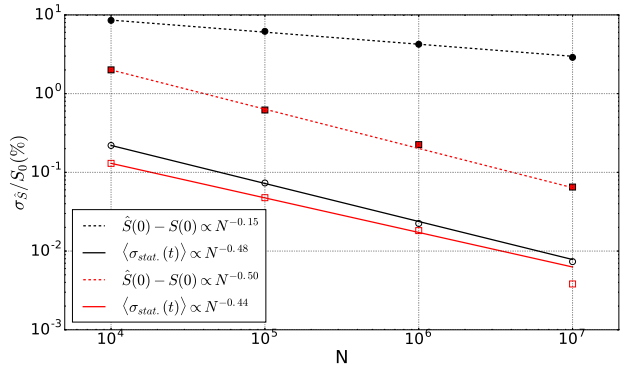}
  \vspace{-0.2cm}
  \caption{$N$-dependence of errors on the entropy estimates. Black
    (red) points represent errors obtained with the initial uniform
    sphere (Plummer sample). Open points are the statistical
    fluctuation obtained as the mean standard deviation of 10
    realizations in each time-step, averaged over all
    time-steps. These statistical fluctuations behave approximately as
    $\sim 1/\sqrt{N}$ (solid lines). Full points represent the error
    in the initial entropy estimate in comparison with the expected
    values, Eqs. \eqref{eq:S_uniform_sph} and \eqref{eq:S_f_E}. The
    deviation of $1/\sqrt{N}$ for the uniform sphere is probably due
    to the non-smoothness in the borders.}
  \label{fig:err_vs_N}
\end{figure}

In Fig.\ref{fig:err_vs_N} we show the $N$-dependence of the
uncertainties $\sigma_{\hat{S}}$ associated with the entropy
estimators for these two different initial conditions. As in
Fig.~\ref{fig:S_plummer}, black dots are obtained for the initial
condition generated from a uniform sphere and the red squares for the
initial Plummer sample. Open points represent the statistical
fluctuation in the entropy estimate at each time-step obtained with 10
different realizations and then averaged over all time-steps for a
fixed $N$. For both initial conditions, we see that these statistical
fluctuations approximately behave as $\sim 1/\sqrt{N}$ (continuous
lines) and this is exactly what is rigorously proven, for the case of
smooth enough distributions -- see \cite{biau2015lectures}.

Full points in Fig. \ref{fig:err_vs_N} represent the systematic error
(bias) in the initial entropy estimate, in comparison to the expected
values given by Eqs. \eqref{eq:S_uniform_sph} and \eqref{eq:S_f_E}. We
observe for the initial uniform sphere that these errors are larger
and have a decay $\propto 1/N^{0.15}$ (dashed black line), slower than
the rigorously proven $1/\sqrt{N}$. This is probably due to the fact
that the uniform sphere distribution is discontinuous in the border,
while, as in the case of the dispersion $\sigma _{\hat{S}}$, the
analytic studies of convergence of the estimators assume smoothness in
all domain -- see \cite{biau2015lectures}. On the other hand, for the
Plummer initial sample, this error with respect to the expected
initial value behaves as $1/\sqrt{N}$ (dashed red line).

\end{document}